\newif\iftwocolumnmode
\twocolumnmodefalse
\newif\ifvcsinfo
\vcsinfofalse
\iftwocolumnmode  % 10pt for 2 column, 12pt for 1 column
  \documentclass[pdftex, 10pt]{iopart}
\else
  \documentclass[pdftex, 12pt]{iopart}
\fi

\usepackage[pdftex]{graphicx}
\usepackage{cite}
\usepackage{siunitx}
\usepackage[pdftex]{color}
\usepackage{comment}
\usepackage{braket}
\usepackage{datetime}
\usepackage[%
  pdftex,
  setpagesize=false,%
  bookmarks=true,%
  bookmarksdepth=tocdepth,%
  bookmarksnumbered=true,%
  colorlinks=true,%
  allcolors=magenta,%
  pdftitle={},%
  pdfsubject={},%
  pdfauthor={Kosuke Mizuno},%
  pdfkeywords={}%
]{hyperref}

\begin{document}

\title[]{Effect of Decoherence for Gate Operations on a Superconducting Bosonic Qubit}
\author{Kosuke Mizuno$^{1\ast}$, Takaaki Takenaka$^1$, Imran Mahboob$^1$, Shiro Saito$^{1\dagger}$}
\address{$^1$NTT Basic Research Laboratory, NTT Corporation, 3-1 Morinosato-Wakamiya, Atsugi, Kanagawa 243-0198, Japan}
\ead{$^\ast$kosuke.mizuno.nv@hco.ntt.co.jp; $^\dagger$shiro.saito.bx@hco.ntt.co.jp}

\ifvcsinfo
  \vspace{10pt}
  \begin{indented}
    \item[]complied at \currenttime~on \today
    \IfFileExists{gitrevision.txt}{\item[]vcs info: \input{gitrevision.txt}}{}
  \end{indented}
\fi

\begin{abstract}
High-quality-factor 3D cavities in superconducting circuits are ideal candidates for bosonic logical qubits as their fidelity is limited only by the low photon loss rate.
However, the transmon qubits that are used to manipulate bosonic qubits result in the emergence of additional relaxation and dephasing channels.
In this work, a numerical study is performed to elucidate the effect of the various loss channels on the performance of logical gates on a bosonic qubit.
A gate error model is developed that encapsulates the loss mechanisms for arbitrary gate operations and predicts experimentally achievable gate errors for bosonic qubits.
The insights gleaned from this study into loss mechanisms suggest more efficient optimization algorithms that could reduce gate errors on bosonic qubits.
\end{abstract}

\vspace{2pc}

%%%%%%%%%%%%%%%%%%%%%%%%%%%%%%%%%%%%%%%%%%%%%%%%%%%%%%%%%%%%%%%%%%%%%%%%%%%%%%%%
\iftwocolumnmode
  \ioptwocol
\fi
\section{Introduction}

Encoding quantum information into a harmonic oscillator mode of a superconducting cavity offers a hardware-efficient approach to quantum computing through the large Hilbert space available to this architecture~\cite{Joshi2021rev}.
The resultant superconducting bosonic qubits have so far yielded unprecedented lifetimes as they are prone to only photon loss~\cite{Ofek2016, Reagor2013APL, Reagor2016PRB}.
More impressively, the photon-loss error can be mitigated via error correcting codes~\cite{Michael2016PRX}, thus offering the possibility of bosonic logical qubits with unprecedented performance.
However, a nonlinear element is required to manipulate these near harmonic qubits and implement various quantum operations with high fidelity~\cite{Ma2021SB} (for instance, state preparation~\cite{Leghtas2013PRA}, logical gate operations~\cite{Heeres2017NC, Reinhold2020NP, Ma2020NPet}, quantum non-demolition (QND) measurements~\cite{Sun2014qnd, Rosenblum2018ftqnd}, error correcting gates~\cite{Hu2019NP, Gertler2021aqec}, and multi-cavity operations~\cite{Wang2016twocat, Gao2018PRX, Gao2019Neswap}).
In superconducting circuits, this requirement can be satisfied by introducing an ancilla transmon qubit that provides a nonlinearity~\cite{Vlastakis2013} with which to manipulate the bosonic qubit by using numerically optimized protocols~\cite{Glaser2015revcat, Heeres2017NC, Hu2019NP}.

Numerical optimization can be used to generate intricate waveforms with which high-fidelity operations can be realized.
However, the transmon introduces not only a nonlinearity but also additional decoherence channels, namely relaxation and dephasing, whose rates are generally higher than the photon loss rate of the cavity.
Although numerical gate implementations on bosonic qubits have been studied, the effects of the underlying decoherence channels on gate errors are challenging to decipher because of the intricate transmon-cavity entangled states that briefly emerge during the optimally controlled gate evolution.
In previous studies, Heeres~\textit{et al}\@.~\cite{Heeres2017NC} reported that the dominant source of gate errors was transmon dephasing, whereas Hu~\textit{et al}\@.~\cite{Hu2019NP} reported that it was transmon relaxation.
Even though these studies employed similar device architectures, their gate errors were limited by different mechanisms.
Therefore, a fundamental question emerges: how do the various decoherence channels affect gate errors on a bosonic qubit?

This work investigates the effect of decoherence, namely transmon relaxation, transmon dephasing, and cavity photon loss, on the errors of numerically optimized gates on bosonic qubits in superconducting circuits.
First, logical gate operations on a bosonic qubit, encoded with the lowest order binomial code~\cite{Michael2016PRX}, were generated by numerical optimization under the assumption of no decoherence.
Then, the gate errors of these operations in the presence of realistic decoherence were evaluated by solving the Lindblad equation.
Next, we developed a gate error model to encapsulate the various decoherence effects on any unitary gate operation.
This model quantitatively describes the contribution of each decoherence channel for the optimized gates.
It yields an approximate lower bound for the achievable gate error on a bosonic qubit.
Specifically, for the Hadamard gate, a \SI{1}{\percent} gate error can be achieved when a transmon with a relaxation time of \SI{100}{\micro\second} and a pure dephasing time of \SI{31}{\micro\second} is coupled to a cavity with a lifetime of \SI{1}{\milli\second}.

Furthermore, the numerical and statistical analyses reveal a qualitative difference between gate errors stemming from transmon relaxation and cavity photon loss.
By comparing errors in repeatedly optimized Hadamard gates, it is found that the contribution from transmon relaxation varies significantly across the gates, whereas the contribution from cavity photon loss exhibits a smaller variance.
Even though the gates each have wholly different waveforms that lead to unique transmon-cavity entangled states as the bosonic qubit evolves, the contribution of the cavity photon loss to the gate error remains insensitive to the underlying gate operation.
The gate error model not only describes the physical mechanism of this insensitivity, but also indicates that it is a feature of cavity modes driven by displacement.
These insights further suggest optimization algorithms that can handle decoherence effects with only modest numerical costs, which could lead to reduced gate errors.

The rest of the paper is organized as follows.
Section~\ref{sec:principle} describes superconducting bosonic qubits and quantum optimal control.
Section~\ref{sec:pre_result} shows errors of numerically optimized gates and the contributions to the error from the cavity photon loss and transmon relaxation.
Section~\ref{sec:model} introduces our gate error model, and Section~\ref{sec:mechanism} describes the properties of the decoherence channels that are revealed with it.
In Section~\ref{sec:approxerror}, the minimum gate error on a binomial bosconic qubit is estimated.
Section~\ref{sec:discussion} discusses optimization algorithms that include decoherence effects with mitigated computational complexity.

%%%%%%%%%%%%%%%%%%%%%%%%%%%%%%%%%%%%%%%%%%%%%%%%%%%%%%%%%%%%%%%%%%%%%%%%%%%%%%%%
\section{Bosonic Qubits and Quantum Optimal Control} 
\label{sec:principle}

\begin{figure*}[htbp]
  \centering
  \includegraphics[width=.75\textwidth]{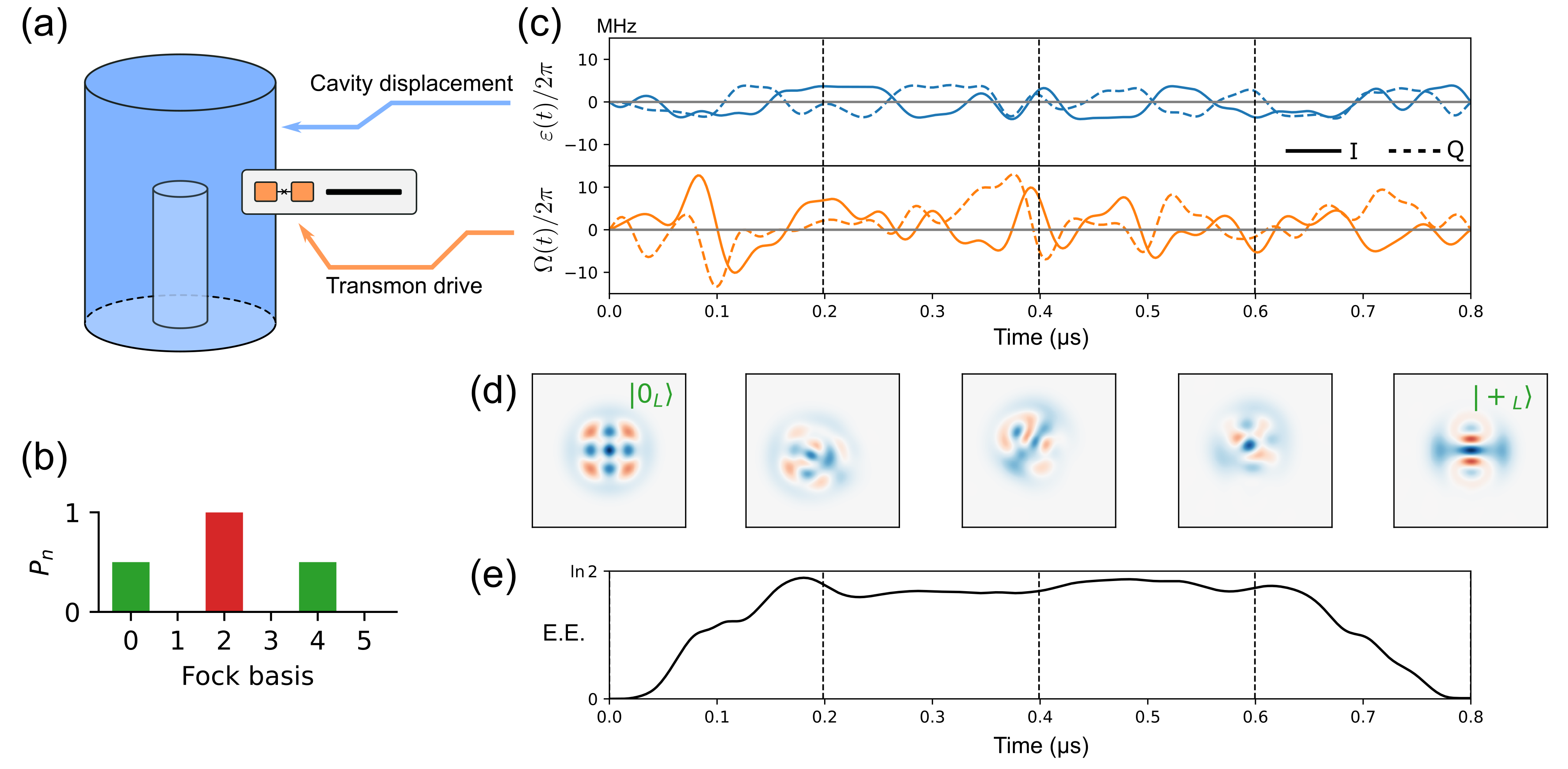}
  \caption{
    (a)
    Bosonic qubit with a superconducting circuit.
    Quantum information is encoded in the cavity (blue) and controlled via the dispersive interaction with the transmon qubit (orange).
    (b)
    Photon distributions $P_n=\left| \langle n|k_L\rangle \right|^2$ of the codewords, $|0_L\rangle$ (green) and $|1_L\rangle$ (red), of the Bin(1,1) code.
    (c)
    Example of control waveforms whose application to the cavity and transmon enables a Hadamard gate to be realized with a gate time of \SI{0.8}{\micro\second}.
    The in- and quadrature-phase components of each control waveform are shown as solid and dashed lines, respectively.
    (d)
    Calculated Wigner functions in the cavity initialized in $|0_L\rangle$ with application of the optimized gate shown in (c).
    Also shown are the intermediate states ($t=0.2$, $0.4$, and $\SI{0.6}{\micro\second}$) as the cavity evolves from the initial to the final state.
    (e)
    Time evolution of entanglement entropy (E.E.) of the system under the gate operation.
  } \label{fig:concept}
\end{figure*}

In a bosonic qubit, quantum information is encoded as a superposition of the Fock states in an oscillator mode in a coaxial cavity (Fig.~\ref{fig:concept}(a), blue)~\cite{Reagor2016PRB}.
Although several bosonic codes~\cite{Gottesman2001, Mirrahimi2014NJP, Albert2018binmod, Grimsmo2020PRX, L.Li2021PRA} have been proposed to protect quantum information from photon loss, binomial codes~\cite{Michael2016PRX} have emerged as an important class due to their compact photon distributions that lead to high-fidelity gate operations.
Here, the lowest order binomial (Bin(1,1)) code is employed, and the logical codewords are defined as
\begin{eqnarray}
 |0_L\rangle = \frac{|0\rangle + |4\rangle}{\sqrt{2}}, \quad |1_L\rangle = |2\rangle.
\end{eqnarray}
The quantum information is protected when an error due to cavity photon loss occurs, as this error can be detected and corrected.
This protection stems from the photon distributions of the codewords having a distinct separation in the Fock basis (Fig.~\ref{fig:concept}(b)), thus enabling the quantum state to be readily identified in the error space, but with the penalty of requiring non-trivial protocols to execute logical gate operations.
Such operations can be implemented by coupling the cavity to an ancilla transmon qubit~\cite{Koch2007PRA}(Fig.~\ref{fig:concept}(a), orange) with a combined system Hamiltonian in a rotating frame given by:
\begin{eqnarray}
  \hat H(t) = \hat H_0 + \hat H_\mathrm{ctrl} (t) , \label{eq:H_sys} \\
  \hat H_0 = \chi \hat a^\dagger \hat a |e\rangle\langle e|, \\
  \hat H_\mathrm{ctrl} (t) = \frac{\Omega_I(t)}{2} \hat\sigma_x + \frac{\Omega_Q(t)}{2} \hat\sigma_y + \varepsilon_I(t) \frac{\hat a + \hat a^\dagger}{2} + \varepsilon_Q(t) \frac{\hat a- \hat a^\dagger}{2i}, \label{eq:H_ctrl}
\end{eqnarray}
where $\hat a$, $\hat\sigma_{x/y}$, and $\chi$ correspond to the annihilation operator of the cavity, the Pauli operators of the transmon, and the dispersive coupling strength, respectively~\cite{Joshi2021rev}.
Eq.(\ref{eq:H_ctrl}) describes the control Hamiltonian for a bosonic qubit, where $\Omega_{I/Q}(t)$ and $\varepsilon_{I/Q}(t)$ correspond to the in- and quadrature-phase amplitudes of the transmon drive (microwave pulses at the transition frequency between $|g\rangle$ and $|e\rangle$ states of the transmon) and the cavity displacement (microwave pulses at the cavity frequency).
These controls constitute a necessary prerequisite to implementing arbitrary gate operations for a bosonic qubit~\cite{Krastanov2015PRA, Heeres2015PRL}.

Non-trivial operations on bosonic qubits are realized by a quantum optimal control based on a numerical optimization with the gradient ascent pulse engineering (GRAPE) algorithm~\cite{Khaneja2005grape}.
The concept of this algorithm is first described for a simple system with a time-dependent Hamiltonian $\hat H(t) = u(t) \hat H_c$, where $u(t)$ is the amplitude and $\hat H_c$ is the Hamiltonian corresponding to the control.
The aim of this optimization is to obtain a waveform $u(t)$ that yields the target unitary operator $\hat U_\mathrm{target}$.
The waveform is parameterized, for example, like a Fourier series $u(t) = \sum_l \left( a_l \cos \omega_l t + b_l \sin \omega_l t \right)$, where $\omega_l$ is the frequency of the $l$-th element.
By dividing the gate time $T_\mathrm{gate}$ into $N$ time steps, the amplitudes are discretized as $u_j = u \left( j\frac{T_\mathrm{gate}}{N} \right)$ for $j=0,1,\ldots, N-1$.
From the Schr\"{o}dinger equation in natural units of $\hbar=1$, the propagator in the $j$-th time step is calculated as $\hat U_j = \exp\left(-i u_j \hat H_c \frac{T_\mathrm{gate}}{N} \right)$.
Thus, the total unitary operator of this control is $\hat U_\mathrm{tot} = \hat U_{N-1} \cdots \hat U_1 \hat U_0$, and the cost function corresponding to the gate error is given by $\Psi = 1 - \left| \frac{1}{d_Q}\tr \left[ \hat U_\mathrm{target}^\dagger \hat U_\mathrm{tot} \right] \right|^2$, where $d_Q=2$ is the dimension of the codespace.
The GRAPE algorithm enables an efficient gradient calculation of the cost function with respect to the control amplitudes $\frac{\partial \Psi}{\partial u_j}$ as well as the gradients with respect to the parameter $a_l$ via the chain rule, namely $\frac{\partial \Psi}{\partial a_l} = \sum_j \frac{\partial \Psi}{\partial u_j} \frac{\partial u_j}{\partial a_l}$.
A gradient-based optimizer minimizes the cost function and yields a parameter set whose total unitary operator emulates the target operator~\cite{deFouquieres2011}.
This method can easily be extended to a system with multiple controls; the optimization for a bosonic qubit is detailed in \ref{ax:grape_detail}.
Note that the cost function includes penalties limiting the waveform amplitude and bandwidth to generate realistic waveforms~\cite{Heeres2017NC}.

The waveforms and resultant errors of an optimized gate critically depend on its initial parameters, which indicates the existence of many local minima that make it challenging to find a globally optimal solution (see \ref{ax:example_waveform}).
This problem can be indirectly addressed by repeating the gate optimizations with multiple random initial parameters.
The gate fidelity $F$ from multiple optimizations can then be evaluated by performing simulations with the Schr\"{o}dinger or Lindblad equations, from which the gate error $r=1-F$ can be calculated (see \ref{ax:fidelity_definition} and \ref{ax:eval_gatefid}).
Four decoherence channels are considered in the simulations reported here: cavity photon loss (with loss rate $\kappa$ and jump operator $\hat a$), transmon relaxation $(1/T_1, \hat \sigma_-)$, pure dephasing of the transmon $(1/T_\phi, \hat \sigma_z/\sqrt{2})$, and thermal excitation of the transmon $(n_\mathrm{th}/T_1, \hat \sigma_+)$.
The Lindblad equation with the aforementioned decoherences can faithfully simulate the system dynamics~\cite{Heeres2017NC}.

We should emphasize that the numerical experiments in this study are composed of two steps.
First, gate optimizations in a closed space are executed with random initial parameters; thus, decoherence is not considered in this step.
Then, the errors of the numerically optimized gates are evaluated by solving the Schr\"{o}dinger and Lindblad equations.

Figure~\ref{fig:concept}(c) shows the waveforms for the Hadamard gate for the Bin(1,1) code, and Fig.~\ref{fig:concept}(d) shows the Wigner functions of the cavity under the gate operation.
The initial state $|g\rangle\otimes|0_L\rangle$ is converted into the final state $|g\rangle\otimes|+_L\rangle$.
Although the cavity and transmon are not entangled before and after the gate, the intermediate states are almost maximally entangled (Fig.~\ref{fig:concept}(e)).
This makes it challenging to use numerical optimization to interpret the effect of decoherence during the gate evolution.
Note that gate operations on bosonic qubits are classified into phase gates and non-phase gates (see \ref{ax:comparison_gate}).
Phase gates (e.g., $Z$ gate and parity mapping operations) generally exhibit high fidelity because they preserve the photon distribution of the cavity states.
On the other hand, non-phase gates (e.g., the $X$ gate, Hadamard gate, error correction gates, and displacement operations) tend to show low fidelity because they change the photon distribution~\cite{Heeres2017NC}.
Most non-phase gates exhibit similar average fidelities; the Hadamard gate investigated in this study is a representative of this gate class.

%%%%%%%%%%%%%%%%%%%%%%%%%%%%%%%%%%%%%%%%%%%%%%%%%%%%%%%%%%%%%%%%%%%%%%%%%%%%%%%%
\section{Gate Errors of Numerically Optimized Gates} 
\label{sec:pre_result}

\begin{figure}[htbp]
  \centering
  \includegraphics[width=.45\textwidth]{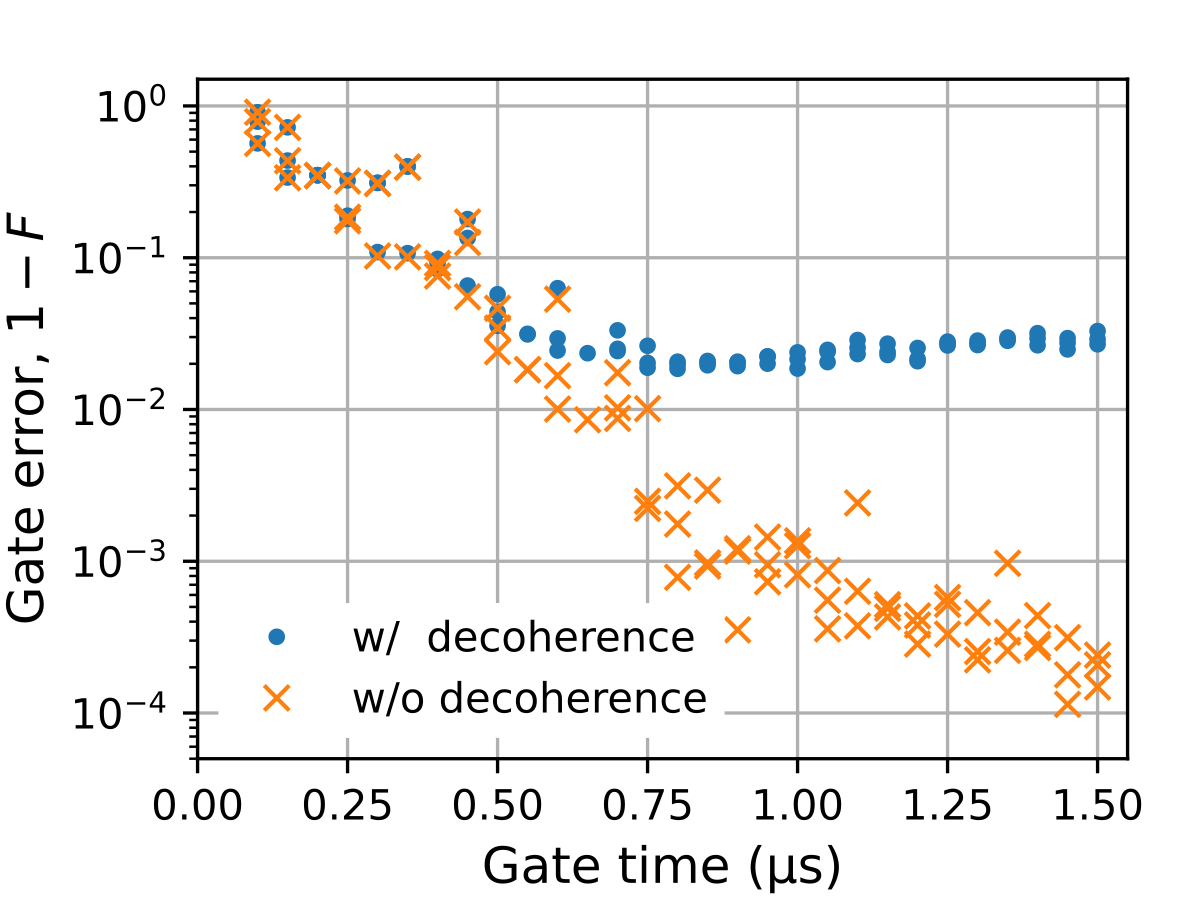}
  \caption{
    Gate errors of optimized Hadamard gates with the Bin(1,1) code.
    Blue dots (orange crosses) correspond to gate errors with (without) docoherence.
    The minimal error is \SI{1.8}{\percent} with realistic decoherence.
    Three optimized gates were generated with random initial parameters for each gate time, yielding the vertical distributions.
  } \label{fig:time_vs_error}
\end{figure}

Figure~\ref{fig:time_vs_error} shows the errors of the optimized Hadamard gates as a function of gate time.
The plots reveal that, although longer gates yield less errors in the absence of decoherence, the error saturates at longer gate times in a more realistic device with finite decoherence rates.
The minimal gate error is \SI{1.8}{\percent} around $T_\mathrm{gate}=\SI{0.8}{\micro\second}$ with $\kappa^{-1}=\SI{1}{\milli\second}$, $T_1=\SI{100}{\micro\second}$, $T_\phi=\SI{25}{\micro\second}$, and $n_\mathrm{th}=0.01$ where the dispersive coupling strength is fixed to $\chi/2\pi=-\SI{2}{\mega\hertz}$ (and is used throughout this study).
Note that state-of-the-art bosonic qubits have superior coherence properties to the above quoted values~\cite{Reinhold2020NP}.

\begin{figure}[htbp]
  \centering
  \includegraphics[width=.45\textwidth]{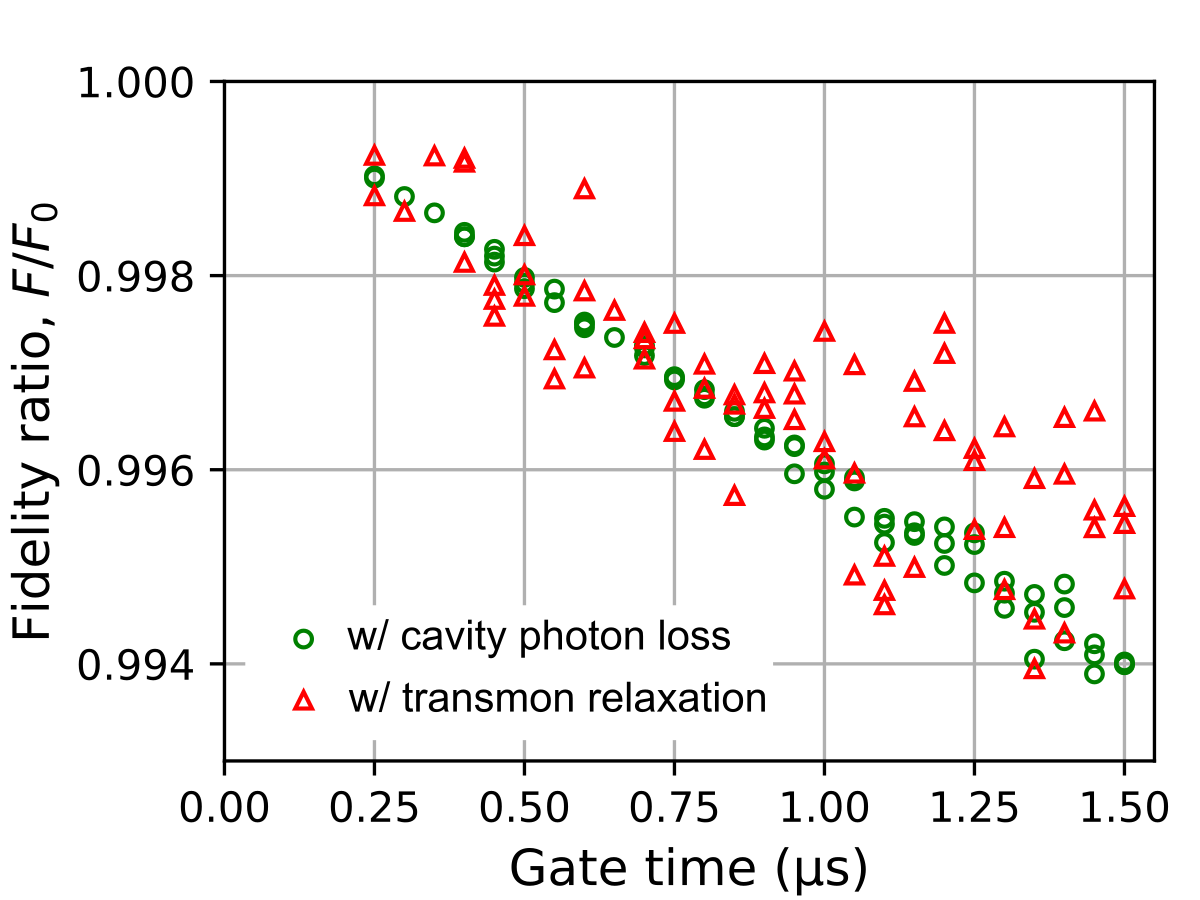}
  \caption{
    Effect of different decoherence channels on optimized Hadmard gates.
    Red triangles correspond to the gate fidelity ratios with only transmon relaxation, and green circles correspond to the ratio with only cavity photon loss.
    The two channels yield similar errors, but the photon loss channel shows a much smaller variation.
    The fidelity ratio is also calculated for three different numerical optimizations at each gate time.
  } \label{fig:randomness}
\end{figure}

The contribution of each decoherence channel to the overall gate error can be revealed by calculating the ratio $F/F_0$ where $F$ is the gate fidelity in the case of a single decoherence channel and $F_0$ is the intrinsic gate fidelity without decoherence.
The red triangles in Fig.~\ref{fig:randomness} correspond to the fidelity ratio in the case of transmon relaxation ($T_1=\SI{100}{\micro\second}$).
These plots exhibit variations stemming from the fact that the cost function for gate optimization has multiple local minima.
For a given gate time, each optimization results in a different waveform that depends on the initial parameter set (see \ref{ax:example_waveform}), which leads to radically different intermediate transmon-cavity entangled states during the gate evolution.
The resultant decoherence is strongly dependent on the details of the intermediate states, which causes a large variation in gate fidelity when the decoherence source is transmon relaxation.
This behavior is consistent with the findings of a previous study on a spin-chain system~\cite{Schulte-Herbruggen2011}.

The fidelity ratio in the case of cavity photon loss (green circles in Fig.~\ref{fig:randomness}) is markedly different from the case of transmon relaxation.
Although a loss rate $\kappa^{-1}=\SI{500}{\micro\second}$ causes a similar magnitude of gate errors as transmon relaxation when averaged at each gate time, the fidelity ratio exhibits a much smaller variation.
It should be emphasized that the variations of the $F/F_0$ ratios for the cavity photon loss and transmon relaxation are radically different for the same optimized gates.
In particular, the effect of the photon loss channel is almost independent of the different intermediate states that emerge during each gate evolution.
In the following section, a gate error model is developed that can describe the features of this contrasting behavior.

%%%%%%%%%%%%%%%%%%%%%%%%%%%%%%%%%%%%%%%%%%%%%%%%%%%%%%%%%%%%%%%%%%%%%%%%%%%%%%%%
\section{Gate Error Model for Arbitrary Gate Operations} 
\label{sec:model}

An error budget analysis that is used in simple protocols like QND measurements based on Ramsey interferometry~\cite{Sun2014qnd, Rosenblum2018ftqnd} can evaluate errors in terms of the product of the gate time, decoherence rate, and impact of a given decoherence.
Although the effect of decoherence is constant in a Ramsey measurement, this is not the case for numerically optimized gate operations.
In what follows, we extend the error budget analysis to handle arbitrary unitary operations.
Our gate error model represents the decoherence-induced gate error as a weighted sum of the decoherence rates of the system, $r^\prime = T_\mathrm{gate} \sum_k \gamma_k s_k$, where $k$ denotes a decoherence channel, $\gamma_k$ the decoherence rate, and $s_k$ the \textit{error susceptibility} (described later).

First, this model is considered in the case of a single decoherence channel with an initial state $|\psi_i\rangle$ and an intermediate state at time $t$ given by $|\psi(t)\rangle$.
The probability of an error in unit time is given by $\gamma_k \cdot p_k(t;\psi_i) = \gamma_k \langle \psi(t) | \hat L_k^\dagger \hat L_k |\psi(t)\rangle$, where $\hat L_k$ corresponds to the jump operator of the decoherence channel $k$, and henceforth we coin $p_k(t;\psi_i)$ as the \textit{error probability} at $t$.
If an error on the state occurs at time $t$, the system falls into $|\psi^\prime(t)\rangle = \frac{\hat L_k |\psi(t)\rangle}{\| \hat L_k |\psi(t)\rangle \|}$~\cite{QCQI}.
The impact of the error can be quantified by the fidelity $l_k(t;\psi_i)$ between the states just before and after, as $l_k(t;\psi_i) = \left| \langle \psi(t) | \psi^\prime(t) \rangle \right|^2$.
Hence, the contribution of decoherence on the gate error in a short time $\Delta t$ is given by the product of the probability and fidelity deterioration:
\begin{eqnarray}
  \Delta r^\prime(t;\psi_i) = \Delta t \cdot \gamma_k \cdot s_k(t;\psi_i), \\
  \text{with}\quad s_k(t;\psi_i) = p_k(t;\psi_i) \cdot \bigg( 1- l_k(t;\psi_i) \bigg),
\end{eqnarray}
where $s_k(t;\psi_i)$ corresponds to the error susceptibility at $t$.
By integrating $\Delta r^\prime(t;\psi_i)$ with respect to time, the gate error induced by decoherence can be determined.
Moreover, the initial state can be any superposition of the logical codewords, so we average it on the logical Bloch sphere ($\int d\psi_i = 1$), yielding:
\begin{eqnarray}
  r^\prime = T_\mathrm{gate} \cdot \gamma_k \cdot s_k, \\
  \text{with} \quad s_k = \int d\psi_i \int_0^{T_\mathrm{gate}} \frac{dt}{T_\mathrm{gate}} s_k(t;\psi_i),  \label{eq:def_s}
\end{eqnarray}
where $s_k$ denotes the error susceptibility of the gate operation.
Next, we generalize the above argument to a case with multiple decoherence channels by expressing the decoherence-induced gate error as a weighted sum of the decoherence rates:
\begin{equation}
  r^\prime = T_\mathrm{gate} \sum_k \gamma_k s_k.  \label{eq:def_rp}
\end{equation}

It should be noted that this approach is premised on (1) a Markovian assumption in which the Lindblad equation describes the system dynamics and (2) weak decoherence, namely $T_\mathrm{gate} \ll 1/\gamma_k$.
A detailed derivation is given in \ref{ax:model}.
By comparing the gate errors estimated by Eq.(\ref{eq:def_rp}) and the Lindblad equation, it can be seen that our analysis quantitatively captures the effects of decoherence on gate errors (see \ref{ax:compare_Lindblad}).

For later convenience, we will introduce \textit{unnormalized fidelity} $l_k^\prime(t;\psi_i)$ at $t$ as $l_k^\prime(t;\psi_i) = p_k(t;\psi_i) \cdot l_k(t;\psi_i) = | \langle \psi(t) |\hat L_k|\psi(t)\rangle |^2 $, which satisfies
\begin{equation}
  s_k(t;\psi_i) = p_k(t;\psi_i) - l_k^\prime(t;\psi_i).
\end{equation}
Explicitly, $s_k(t;\psi_i)$ and $p_k(t;\psi_i)$ denote the contribution of decoherence to the gate error and the probability of decoherence, respectively.
Thus, $l_k^\prime(t;\psi_i)$ indicates how tolerant the state is to decoherence.
Moreover, $p_k$ and $l_k^\prime$ denote the average values of $p_k(t;\psi_i)$ and $l_k^\prime(t;\psi_i)$ in analogy to Eq.(\ref{eq:def_s}); they satisfy $s_k = p_k - l_k^\prime$.

%%%%%%%%%%%%%%%%%%%%%%%%%%%%%%%%%%%%%%%%%%%%%%%%%%%%%%%%%%%%%%%%%%%%%%%%%%%%%%%%
\section{Decoherence in Bosonic Qubits during Gate Operations} \label{sec:mechanism}

\begin{figure*}[htbp]
  \centering
  \includegraphics[width=.9\textwidth]{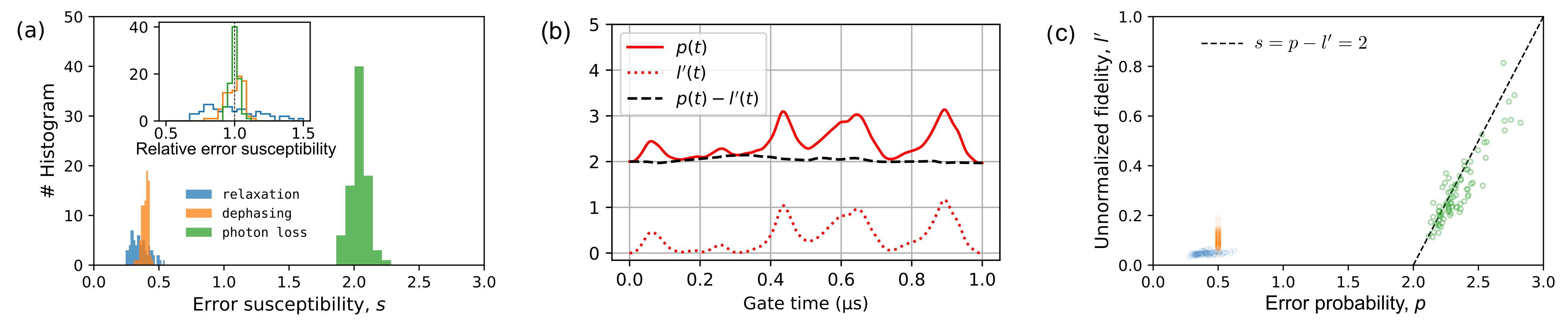}
  \caption{
    (a)
    Histogram of error susceptibility in each decoherence channel (blue: transmon relaxation, orange: transmon dephasing, green: cavity photon loss) evaluated for 100 optimized gates.
    Inset: histograms of error susceptibilities relative to their averages, $s_k / \mathrm{Ave}[s_k]$.
    Same bins are used for the main and inset plots.
    (b)
    Example of $p_\mathrm{loss}(t;\psi_i)$, $l^\prime_\mathrm{loss}(t;\psi_i)$, and $s_\mathrm{loss}(t;\psi_i) = p_\mathrm{loss}(t;\psi_i) - l^\prime_\mathrm{loss}(t;\psi_i)$ for cavity photon loss error.
    (c)
    Scatter plots of error probability $p_k$ and unnormalized fidelity $l^\prime_k$.
    Each dot corresponds to a value calculated from a single optimization result.
    The dashed line is a guide for the eye.
    For the photon loss channel, $p_\mathrm{loss}$ and $l^\prime_\mathrm{loss}$ clearly correlate.
    The error susceptibility $s_\mathrm{loss}$ is almost 2, i.e., the mean photon number of the codewords, and this leads to a small variation in the photon loss as shown in Fig.~\ref{fig:randomness}.
  } \label{fig:susceptibility_analysis}
\end{figure*}

To evaluate the effect of decoherence on a bosonic qubit, the error susceptibilities were calculated for the cavity photon loss, transmon relaxation, and pure dephasing of transmon for 100 optimized gates with $T_\mathrm{gate} = \SI{1}{\micro\second}$.
Thermal excitation of the transmon was ignored since it contributes approximately \SI{1}{\percent} to the transmon relaxation.
Figure~\ref{fig:susceptibility_analysis}(a) shows histograms of the error susceptibility $s_k$.
The relative standard deviation (RSD), defined as the ratio of the standard deviation to the average $\mathrm{Std}[s_k] / \mathrm{Ave}[s_k]$, is 0.033 for the cavity photon loss, 0.068 for transmon dephasing, and 0.208 for transmon relaxation (Table~\ref{tab:s} and Fig.~\ref{fig:susceptibility_analysis}(a, inset)).
The average error susceptibilities for transmon relaxation and dephasing are close to $1/3$, which is reminiscent of the average idle gate error in a two-level system $r^\prime_I = T_\mathrm{gate} \cdot \left(\frac{1}{3 T_1} + \frac{1}{3 T_\phi} \right)$ (Ref.~\cite{Omalley2015PRAppl}, see also \ref{ax:error_transmon}).
However, their standard deviations are different ($\mathrm{Std}[s_\mathrm{dep}] < \mathrm{Std}[s_\mathrm{relax}]$), and the RSD for transmon dephasing is small.
On the other hand, the average error susceptibility for the cavity photon loss is larger than the average for the transmon relaxation ($\mathrm{Ave}[s_\mathrm{loss}] > \mathrm{Ave}[s_\mathrm{relax}]$), while their standard deviations are similar ($\mathrm{Std}[s_\mathrm{loss}] \simeq \mathrm{Std}[s_\mathrm{relax}]$).
The corresponding RSD for the cavity photon loss is much smaller than the RSD for transmon relaxation, thus confirming the observations in Fig.~\ref{fig:randomness}.
Although the contrasting behaviors of the transmon relaxation and cavity photon loss can be described in terms of the RSD of their error susceptibilities, two questions arise: (1) why is $\mathrm{Ave}[s_\mathrm{loss}]$ larger than the average of the transmon relaxation despite their standard deviations being similar?
(2) Why is $\mathrm{Std}[s_\mathrm{dep}]$ smaller than the standard deviation of the transmon relaxation despite their averages being similar?

\Table{\label{tab:s}
  Average, standard deviation, and relative standard deviation (RSD) of error susceptibilities for each decoherence channel in Fig.~\ref{fig:susceptibility_analysis}(a).
}
\br
Decoherence channel & Average & Standard deviation & RSD \\
\mr
Transmon relaxation & 0.363   & 0.075  & 0.208   \\
Transmon dephasing  & 0.396   & 0.027  & 0.068   \\
Cavity photon loss  & 2.038   & 0.069  & 0.033   \\
\br
\endTable

Interestingly, the average error susceptibility of cavity photon loss of 2.038 is close to the mean photon number of the underlying codewords $\bar{n}=2$.
To confirm this observation, error susceptibilities were calculated for other bosonic codes with higher mean photon numbers (4-leg cat code with $\bar{n} \simeq 3$ and the Bin(2,2) code with $\bar{n}=4.5$, see \ref{ax:code_definition}).
In these encodings, the error susceptibilities for photon loss are also distributed around their mean photon numbers (see \ref{ax:approxerror_othercode}).
To account for this similarity, we consider the cavity mode undergoes a displacement, namely $\epsilon_I$ and $\epsilon_Q$ in Eq.(\ref{eq:H_ctrl}) during the gate operation.
This displacement increases the mean photon number, which intuitively suggests a larger gate error induced by photon loss.
However, the displacement concurrently broadens the photon distribution, making it less sensitive to photon loss.
For example, the first Fock state is completely destroyed by photon loss: $\hat a|1\rangle = |0\rangle$, but the coherent state is not affected: $\hat a |\alpha\rangle \propto |\alpha\rangle$.
To verify this explanation, Fig.~\ref{fig:susceptibility_analysis}(b) shows an example of $p_\mathrm{loss}(t;\psi_i)$, $l^\prime_\mathrm{loss}(t;\psi_i)$, and $s_\mathrm{loss}(t;\psi_i)$ during an optimized Hadamard gate operation on the Bin(1,1) code with $|\psi_i\rangle = |g\rangle\otimes\frac{1}{\sqrt{2}}(|0_L\rangle+|1_L\rangle)$.
Note that $p_\mathrm{loss}(t;\psi_i) = \langle\psi(t)|\hat a^\dagger \hat a|\psi(t)\rangle$ corresponds to the instantaneous mean photon number at $t$, and $l^\prime_\mathrm{loss}(t;\psi_i)$ corresponds to how tolerant the intermediate state is to photon loss.
The gate operation causes $p_\mathrm{loss}(t;\psi_i)$ to increase from its initial value of $\bar{n}$ because the cavity undergoes displacement.
Interestingly, the temporal dynamics of $l^\prime_\mathrm{loss}(t;\psi_i)$ are similar to $p_\mathrm{loss}(t;\psi_i)$ thus canceling each other, and consequently, $s_\mathrm{loss}(t;\psi_i)$ remains close to $\bar{n}=2$.
For a simple system, a cavity without an ancilla, the dynamics of $p_\mathrm{loss}(t;\psi_i)$ and $l^\prime_\mathrm{loss}(t;\psi_i)$ exactly coincide.
Due to the fact that $p_\mathrm{loss}(0;\psi_i)=\bar{n}$ and $l_\mathrm{loss}(0;\psi_i)=0$, the relationship $s_\mathrm{loss}(t;\psi_i) = \bar{n}$ is thus satisfied, making this a key feature of cavity modes encoded with bosonic codes (see \ref{ax:toymodel}).

As an illustration of relationship between error probability and unnormalized fidelity, Fig.~\ref{fig:susceptibility_analysis}(c) shows $p_k$ and $l_k^\prime$ for the optimized gates used in Fig.~\ref{fig:susceptibility_analysis}(a).
Note that $p_k$ and $l_k^\prime$ were averaged over time and initial states for each gate.
It shows that the cavity photon loss exhibits a clear correlation with $p_\mathrm{loss}$ and $l^\prime_\mathrm{loss}$ in line with the explanation above.
Note that the green dots in the figure are scattered around the dashed line indicating $s=\bar{n}$.
This distribution emerges from the dispersive interaction and the resultant entanglement between the cavity and the transmon.
These observations were further verified in two additional cases: Hadamard gates with stronger displacement and error correction gates (see \ref{ax:comparison_strong_displacement} and \ref{ax:comparison_QEC}).

The above findings indicate that even large displacements do not increase the gate error due to photon loss during gate operations.
This insight is important because a larger displacement allows for a shorter gate, leading to smaller errors~\cite{Eickbusch2021ecd}.
Indeed, it explains the observations in Ref.~\cite{Eickbusch2021ecd}, in which faster operations were demonstrated by utilizing large displacements that resulted in mean photon numbers up to 2500.
In spite of this, the cavity photon loss still inflicts smaller errors on the gate operations compared with the transmon decoherence channels.

In contrast to the case of the cavity photon loss, the error susceptibility for transmon dephasing is determined only by the unnormalized fidelity: $s_\mathrm{dep} = 1/2 - l^\prime_\mathrm{dep}$.
Note that $p_\mathrm{dep} = \frac{1}{2}\langle\psi(t)|\hat\sigma_z^\dagger\hat\sigma_z|\psi(t)\rangle = 1/2$.
Moreover, the error susceptibility for transmon relaxation is approximately determined by the error probability: $s_\mathrm{relax} \simeq p_\mathrm{relax}$.
From $l^\prime_\mathrm{dep}(t;\psi_i) = \frac{1}{2}\langle\psi(t)|\hat\sigma_z|\psi(t)\rangle^2$ and $p_\mathrm{relax}(t;\psi_i) = \langle\psi(t)|\hat\sigma_-^\dagger \hat\sigma_-|\psi(t)\rangle = \frac{1}{2} + \frac{1}{2}\langle\psi(t)|\hat\sigma_z|\psi(t)\rangle$, the standard deviation of the error susceptibility for transmon dephasing is roughly of the order $\langle\hat\sigma_z\rangle^2$, whereas for relaxation it is $\langle\hat\sigma_z\rangle$.
Since the expectation value of $\hat\sigma_z$ spans $[-1, 1]$, the error susceptibility for dephasing has a smaller variance than that for relaxation, as shown in Fig.~\ref{fig:susceptibility_analysis}(a) (see also \ref{ax:std_s_transmon}).

%%%%%%%%%%%%%%%%%%%%%%%%%%%%%%%%%%%%%%%%%%%%%%%%%%%%%%%%%%%%%%%%%%%%%%%%%%%%%%%%
\section{Discussion}

\subsection{Bounding of Gate Errors for Bin(1,1) Bosonic Qubits} 
\label{sec:approxerror}

In this section, we will discuss the experimentally achievable gate errors.
Gate errors can be decomposed into intrinsic and decoherence contributions: $r= r_0+ r^\prime$.
Even without decoherence, an optimized gate has a non-zero error, which we term intrinsic error $r_0$ that can be evaluated by solving the Schr\"{o}dinger equation.

First, the results of Secion~\ref{sec:mechanism} allows us to estimate the lower bound on decoherence-induced gate errors $r^\prime$.
It can be approximated with Eq.(\ref{eq:def_rp}):
\begin{equation}
  r^\prime \ge T_\mathrm{gate} \cdot \bigg(  \frac{1}{T_1} \cdot 0.25 + \frac{1}{T_\phi} \cdot 0.31 + \kappa \cdot 0.94 \bar{n}  \bigg), \label{eq:approx_1us}
\end{equation}
where the minimum error susceptibility for each channel has been utilized for the prefactors.
The prefactor for the photon loss term is divided by the mean photon number.
Substituting $T_\mathrm{gate}=\SI{1}{\micro\second}$, $\bar{n}=2$, $\kappa^{-1}=\SI{1}{\milli\second}$, $T_1=\SI{100}{\micro\second}$, and $T_\phi=\SI{25}{\micro\second}$ into Eq.(\ref{eq:approx_1us}) yields a gate error of \SI{1.68}{\percent}.
The minimum gate error of the optimized gates used in Fig.~\ref{fig:susceptibility_analysis} was \SI{1.76}{\percent}, and thus we can approximate the lower bound by Eq.(\ref{eq:approx_1us}).
Note that the error susceptibility exhibits only a weak dependence on the gate time, and it effectively can be ignored (see \ref{ax:comparison_gatetime}).

Secondly, in the above case with $T_\mathrm{gate} = \SI{1}{\micro\second}$, the intrinsic error is as small as $r_0 \simeq 10^{-3}$.
However, in the case of short gates, the intrinsic error increases and cannot be ignored.
A detailed investigation for the sources of the intrinsic gate error is beyond the scope of this paper, but a numerical analysis reveals its gate-time dependence can be empirically approximated as $r_0 \sim \exp(-11.05 \cdot T_\mathrm{gate}[\si{\micro\second}])$ (see \ref{ax:approxerror_othercode}).
While further study is needed to arrive at an explanation for this bound, the total achievable gate error can then be approximated as:
\begin{equation}
  r \ge e^{- 11.05 \cdot T_\mathrm{gate}[\si{\micro\second}] } + T_\mathrm{gate} \cdot \bigg(  \frac{1}{T_1} \cdot 0.25 + \frac{1}{T_\phi} \cdot 0.31 + \kappa \cdot 0.94 \bar{n}  \bigg), \label{eq:approx}
\end{equation}
for a Hadamard gate on a Bin(1,1)-encoded bosonic qubit in a superconducting circuit.
Note that in order to approximate the intrinsic error, we modified optimization penalties in the cost function to allow stronger and wider bandwidth waveforms (see \ref{ax:grape_detail}).
With such conditions, the speed of system evolutions under optimized gates are limited mainly due to the dispersive coupling.

\begin{figure}[htbp]
  \centering
  \includegraphics[width=.45\textwidth]{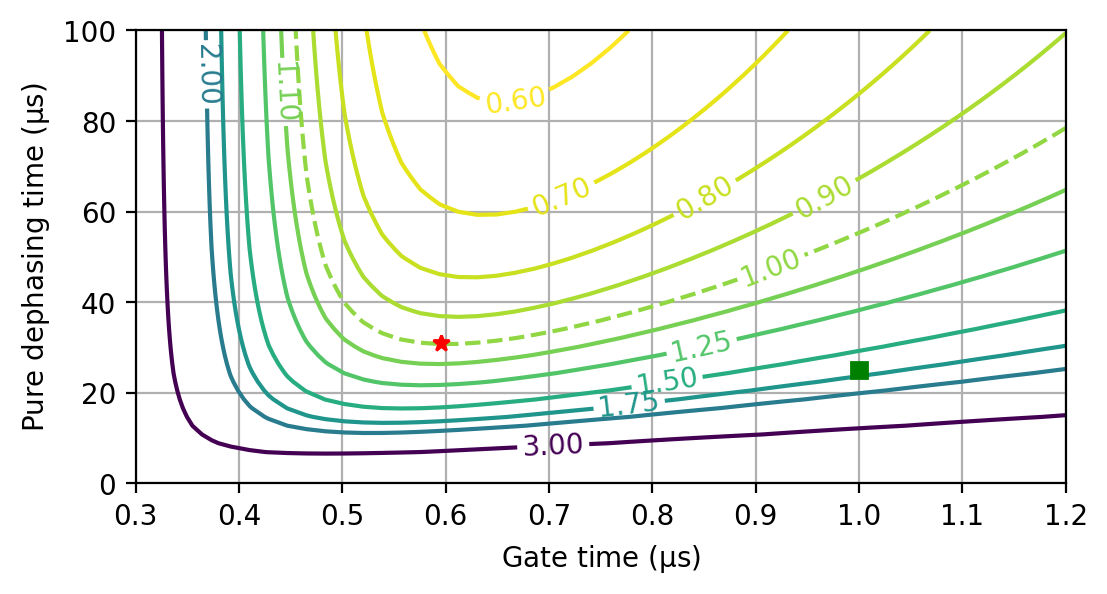}
  \caption{
    Approximate error rate in unit of percent determined from Eq.(\ref{eq:approx}) as a function of gate time $T_\mathrm{gate}$ and pure dephasing time for the transmon $T_\phi$.
    The green square and red star correspond to $(T_\mathrm{gate}, T_\phi)=(\SI{1}{\micro\second}, \SI{25}{\micro\second})$ and $(\SI{595}{\nano\second}, \SI{31}{\micro\second})$.
    The other parameters are $\bar{n}=2$, $\kappa^{-1}=\SI{1}{\milli\second}$, and $T_1=\SI{100}{\micro\second}$.
  } \label{fig:ax_approxerror}
\end{figure}

One approach to fault-tolerant quantum computing would be to implement surface code~\cite{Fowler2012PRA} via bosonic qubits.
Thus, an intriguing question emerges: how long should the coherence time of a bosonic qubit be in order to achieve the surface code threshold at \SI{1}{\percent}~\cite{Barends2014Nature}?
According to Eq.(\ref{eq:approx}) and for the coherence times used in the simulations in Fig.~\ref{fig:time_vs_error}, the pure dephasing channel of the transmon is the dominant source of gate errors.
By sweeping the gate time and pure dephasing time, a \SI{1}{\percent} gate error becomes possible if $T_\phi \geq \SI{31}{\micro\second}$ (Fig.~\ref{fig:ax_approxerror}).
In addition, \SI{0.9}{\percent}, \SI{0.8}{\percent}, \SI{0.7}{\percent}, and \SI{0.6}{\percent} gate errors require even longer pure dephasing times of $T_\phi=\SI{37}{\micro\second}$, $\SI{46}{\micro\second}$, $\SI{60}{\micro\second}$, and $\SI{85}{\micro\second}$, respectively.
Although increasing the dispersive coupling strength enables faster operations and thus reduces gate errors, it comes at the cost of unwanted nonlinearities, such as the self-Kerr interaction~\cite{nigg_black-box_2012}.

%%%%%%%%%%%%%%%%%%%%%%%%%%%%%%%%%%%%%%%%%%%%%%%%%%%%%%%%%%%%%%%%%%%%%%%%%%%%%%%%
\subsection{Optimization Algorithms with Decoherence Effects} \label{sec:discussion}

The results of the numerical evaluation of the error susceptibility naturally suggests two modifications to the optimization algorithm that will lower the gate errors by including the effects of decoherence in the optimization.
Here, our gate synthesis is composed of two steps: gate optimization in a closed system (without decoherence) and a realistic evaluation in an open system (with decoherence), as the latter requires greater computational complexity.
Although previous studies have shown that open-system optimizations yield smaller gate errors~\cite{Rebentrost2009PRL, Schulte-Herbruggen2011, Goerz2014NJP}, in practice this is difficult to achieve for bosonic qubits owing to their large system dimensions.
For example, the optimization in this study with the Schr\"{o}dinger equation was conducted with $60$ dimensions (i.e., $2 \times 30$ for transmon and cavity, respectively).
Replacing it with the Lindblad equation would naively require $60^2=3600$ dimensions, making it computationally exorbitant.

The insight offered by this study is that the effect of the cavity photon loss on the gate errors can be predicted from the cavity lifetime, the mean photon number of the code, and the gate time.
Consequently, we propose a hybrid approach consisting of the Lindblad equation for the transmon and the Schr\"{o}dinger equation for the cavity,
which enables optimization that integrates the decoherence channels of bosonic qubits.
Such a half-open system would require only $2^2 \times 30 = 120$ dimensions, a significant reduction in computational complexity.

Although this half-open approach is tailored for bosonic qubits, the results of this study also support a closed approach that can suppress the numerical cost in general quantum systems.
Specifically, if the error susceptibility is directly included in the cost function, this will permit gate optimizations with decoherence effects requiring only $60$ dimensions.
Unfortunately, the GRAPE algorithm is unable to calculate error susceptibility as quantities that depend on intermediate states require a computationally complex gradient calculation.
However, a recent study has shown that a backpropagation-based algorithm can enable an efficient gradient calculation of such quantities~\cite{Leung2017PRAautograd}.
The error susceptibility can be included in the backpropagation network, and thus, a quantum optimal control with decoherence effects can be realized for any quantum system.

%%%%%%%%%%%%%%%%%%%%%%%%%%%%%%%%%%%%%%%%%%%%%%%%%%%%%%%%%%%%%%%%%%%%%%%%%%%%%%%%%
%\subsection{Numerical Study} \label{sec:limitation}
%Finally, let us briefly comment on the numerical and statistical nature of this study.
%In general, gate operations with smaller error susceptibility yield lower gate errors.
%Thus, a theoretical lower bound for the error susceptibility is essential for the realization of high-fidelity gate operations in bosonic qubits.
%The results of this study offer a numerical answer to this challenging question, but a fundamental understanding of its origin remains an open theoretical inquiry and will be the subject of future work.

%%%%%%%%%%%%%%%%%%%%%%%%%%%%%%%%%%%%%%%%%%%%%%%%%%%%%%%%%%%%%%%%%%%%%%%%%%%%%%%%
\section{Conclusion} 
\label{sec:conclusion}

A bosonic qubit in a superconducting circuit with optimally controlled gates offers an attractive approach to quantum computing.
However, a detailed numerical model incorporating all the decoherence channels that limit the gate fidelities in such a system has remained elusive.
In this work, the performance of a bosonic qubit was modeled with numerical simulations, which yielded an analytical formulation that encapsulates decoherence effects from the cavity and the ancilla transmon constituting the superconducting circuit.
The analysis permitted a quantitative evaluation of decoherence-induced gate errors on the bosonic qubit.
By approximating the achievable gate error of the Hadamard gate on the Bin(1,1)-encoded bosonic qubit, it was found that a gate error of \SI{1}{\percent} is possible by reducing pure dephasing in the transmon.
Moreover, the contribution of the cavity photon loss to the gate error is insensitive to the details of the gate operation.
This insight and our error model suggest alternative optimization algorithms that can handle decoherence at modest numerical costs.

%%%%%%%%%%%%%%%%%%%%%%%%%%%%%%%%%%%%%%%%%%%%%%%%%%%%%%%%%%%%%%%%%%%%%%%%%%%%%%%%
% ACKNOWKEDGEMENTS and REFERENCES
\ack
\addcontentsline{toc}{section}{Acknowledgments}

We thank Dr.~William J.~Munro for fruitful discussions and providing calculation resources.
We thank Dr.~Yuichiro Matsuzaki for fruitful discussions.
This work was supported by JST Moonshot R\&D, Grant Number JPMJMS2067.

\subsection*{Author Contributions}

K.M.~prepared the program codes, conducted the numerical experiments, conceived the gate error model, analyzed all of the data, and wrote the manuscript.
T.T., I.M., and S.S.~contributed to the discussions.
All authors contributed to refining the manuscript.

\subsection*{Data Availability}

The data that support the findings of this study are available from the corresponding author upon reasonable request.

\section*{References}
\addcontentsline{toc}{section}{References}
\bibliography{main}
\bibliographystyle{iopart-num}

%%%%%%%%%%%%%%%%%%%%%%%%%%%%%%%%%%%%%%%%%%%%%%%%%%%%%%%%%%%%%%%%%%%%%%%%%%%%%%%%
\appendix

%%%%%%%%%%%%%%%%%%%%%%%%%%%%%%%%%%%%%%%%%%%%%%%%%%%%%%%%%%%%%%%%%%%%%%%%%%%%%%%%
\section{Definitions}

\subsection{Fidelities and Gate Errors} 
\label{ax:fidelity_definition}

We calculate the fidelity between two quantum states $\hat\rho$ and $\hat\sigma$ by using the following equation:
\begin{equation}
  F(\rho, \sigma) = \left(  \tr \sqrt{\sqrt{\hat\rho} \, \hat\sigma \sqrt{\hat\rho}} \right)^2.
\end{equation}
The average gate fidelity $F$ between $\hat U$ and a quantum process $\mathcal{E}(\hat\rho)$ is calculated as follows:
\begin{eqnarray}
  F = \int d\psi \langle \psi| \hat U^\dagger \mathcal{E}\left( |\psi\rangle\langle\psi| \right) \hat U |\psi\rangle, \label{eq:ax_def_fidelity}
\end{eqnarray}
where $\int d\psi = 1$ is an average on a logical Bloch sphere.
The gate error $r$ of $\hat U$ is evaluated as $r = 1 - F$.

\subsection{Codewords and Operations} 
\label{ax:code_definition}

Let $|0_L\rangle$ and $|1_L\rangle$ be logical codewords of a bosonic code.
After a photon loss error happens, the codeword jumps to a corresponding error word $|k_E\rangle = \hat a |k_L\rangle / \|\hat a |k_L\rangle\|$ for $k=\lbrace 0,1 \rbrace$.
The Bin(1,1) code (mean photon number $\bar n = 2$) corresponds to 
\begin{eqnarray}
 |0_L\rangle = \frac{\ket{0} + \ket{4}}{\sqrt{2}}, \quad |1_L\rangle = |2\rangle, \\
 |0_E\rangle = |3\rangle, \quad |1_E\rangle = |1\rangle.
\end{eqnarray}
The Bin(2,2) code ($\bar n = 4.5$) corresponds to 
\begin{eqnarray}
 |0_L\rangle = \frac{|0\rangle + \sqrt{3} |6\rangle}{2}, \quad |1_L\rangle = \frac{\sqrt{3} |3\rangle + |9\rangle}{2}, \\
 |0_E\rangle = |5\rangle, \quad |1_E\rangle = \frac{|2\rangle + |8\rangle}{\sqrt{2}}.
\end{eqnarray}
The 4-leg cat code corresponds to 
\begin{eqnarray}
 |0_L\rangle &= \frac{1}{\mathcal{N}_0} \bigg( |\alpha\rangle + |-\alpha\rangle  + |i\alpha\rangle + |-i\alpha\rangle \bigg), \\
 |1_L\rangle &= \frac{1}{\mathcal{N}_1} \bigg( |\alpha\rangle + |-\alpha\rangle  - |i\alpha\rangle - |-i\alpha\rangle \bigg),
\end{eqnarray}
where $\mathcal{N}_{0/1}$ are normalization factors to ensure $\langle 0_L|0_L\rangle=\langle 1_L|1_L\rangle = 1$.
The mean photon number is approximately $\bar{n} \simeq |\alpha|^2$.

A gate unitary corresponding to a logical gate operation is a Pauli operator on the logical codewords, such as:
\begin{eqnarray}
 \hat X_L &= \hat 1_t \otimes \bigg( |0_L\rangle\langle 1_L| + |1_L\rangle\langle 0_L| \bigg), \\
 \hat Z_L &= \hat 1_t \otimes \bigg( |0_L\rangle\langle 0_L| - |1_L\rangle\langle 1_L| \bigg),
\end{eqnarray}
where $\hat 1_t$ is the identity operator of the transmon.
Note that the transmon remains in the ground state before and after the operation.
An error recovery gate $\hat U_\mathrm{QEC}$ is a unitary operator obeying the following equation:
\begin{eqnarray}
 \hat U_\mathrm{QEC} \, |g\rangle \otimes \bigg( c_0 |0_E\rangle + c_1 |1_E\rangle \bigg) = |g\rangle \otimes \bigg ( c_0 |0_L\rangle + c_1 |1_L\rangle \bigg ), \label{eq:ax_U_QEC}
\end{eqnarray}
where $c_0$ and $c_1$ are complex numbers satisfying $|c_0|^2 + |c_1|^2 = 1$.

%%%%%%%%%%%%%%%%%%%%%%%%%%%%%%%%%%%%%%%%%%%%%%%%%%%%%%%%%%%%%%%%%%%%%%%%%%%%%%%%
\section{Details of the Calculation}

\subsection{Gate Optimization} 
\label{ax:grape_detail}

The system Hamiltonian $\hat H$ is 
\begin{eqnarray}
 \hat H = \hat H_0 + \hat H_\mathrm{ctrl}, \\
 \hat H_0 = \chi \hat a^\dagger \hat a |e\rangle\langle e|, \\
 \hat H_\mathrm{ctrl} = \frac{\Omega_I(t)}{2} \hat\sigma_x + \frac{\Omega_Q(t)}{2} \hat\sigma_y + \varepsilon_I(t) \frac{\hat a + \hat a^\dagger}{2} + \varepsilon_Q(t) \frac{\hat a- \hat a^\dagger}{2i}.
\end{eqnarray}
Higher order interactions such as the self-Kerr interaction were ignored in this study.
The control Hamiltonian $\hat H_\mathrm{ctrl}$ has four time-dependent terms: (the transmon drive and the cavity displacement) $\times$ (in- and quadrature-phase components).
The four control amplitudes $\Omega_{I/Q}(t)$ and $\epsilon_{I/Q}(t)$ were optimized.
The gate time $T_\mathrm{gate}$ is discretized into $N$ time steps.
The system Hamiltonian at the $j$-th time step can be written in a generalized form,
\begin{equation}
 \hat H_j = \hat H_0 + \sum_k u_{kj} \hat H_\mathrm{k},
\end{equation}
where $j$ and $k$ are labels for time steps and controls, respectively.
The total unitary is given by
\begin{eqnarray}
 \hat U_j &= \exp(-i \hat H_j \Delta t) \\
 \hat U_\mathrm{tot} &= \hat U_{N-1} \dots \hat U_1 \hat U_0.
\end{eqnarray}
Note that we use natural units of $\hbar = 1$.

The cost function of the optimization includes a gate error and amplitude penalties.
The gate error can be calculated in terms of the Hilbert--Schmidt distance:
\begin{equation}
 \Psi_1 = 1 - \frac{1}{{d_Q}^2} \bigg| \mathrm{tr}[ \hat P_Q \hat U_\mathrm{targ}^\dagger \hat P_Q \hat U_\mathrm{tot} ] \bigg|^2,
\end{equation}
where $d_Q=2$ is the dimension of the code, and $\hat P_Q = |g\rangle\langle g| \otimes \left( |0_L\rangle\langle 0_L| + |1_L\rangle\langle 1_L| \right)$ is the projection operator to the codespace~\cite{Motzoi2011PRA}.
Multiplying the projection operations prevents leakage out of the codespace.
The global phase of this gate was ignored in the optimization.
The GRAPE algorithm~\cite{Khaneja2005grape} enables fast evaluations to be made of the cost gradients of the gate error, $\partial \Psi_1 / \partial u_{kj}$.

The number of dimensions in the gate optimization was 60 (2 for the transmon $\times$ 30 for the cavity).
Note that ignoring the second excited state of the transmon has negligible effect.
Although, in principle, a cavity mode is a harmonic oscillator of infinite dimension, the cavity dimensions can be truncated for numerical computations.
Here, amplitude penalties are given to suppress occupations of higher photon states during the gate to avoid truncation-related effects.
A nonlinear amplitude penalty was used instead of a simple quadratic penalty:
\begin{equation}
 \Psi_2 = \frac{1}{N} \sum_j \exp\left[\left(\frac{u_{kj}}{u_{k}^\mathrm{(max)}}\right)^4\right].
\end{equation}
This penalty softly clips the amplitudes at $u_{k}^\mathrm{(max)}$.

The time-dependent amplitudes were parameterized with $2M+1$ degrees of freedom for each control in the frequency domain:
\begin{eqnarray}
 u_{kj} = c_0 + \sum_{l=1}^M \left( a_{kl} \cos(2\pi f_l t_j) + b_{kl} \sin(2\pi f_l t_j) \right), \\
 \mathrm{with} \quad f_l = l \frac{f_\mathrm{max}}{M},
\end{eqnarray}
where $f_\mathrm{max}$ corresponds to the maximum bandwidth.
$M$ was set to a number large enough to ensure a sufficient degree of freedom of $u_{kj}$ according to signal processing theory~\cite{Slepian1978}.

To ensure that the waveforms would be within the desired bandwidth, amplitude penalties were added at time boundaries:
\begin{eqnarray}
 \Psi_3 = |u_{k0}|^2 + |u_{k\,N-1}|^2.
\end{eqnarray}
This penalty makes the waveforms start and end smoothly.

The cost function is a weighted sum of the above penalties $\Psi = \sum_i c_i \Psi_i$, and the coefficients $c_i$ were determined empirically.
The total number of optimization parameters is $4(2M+1)$, typically being a few hundred.
We started an optimization with random initial parameters and sequentially updated the parameters to minimize the cost function with a limited-memory Broyden--Fletcher--Goldfarb--Shanno algorithm~\cite{scipy}.

For the optimizations described in most of this study, $f_\mathrm{max}=\SI{30}{\mega\hertz}$, $\Delta t=\SI{2}{\nano\second}$, $u_\mathrm{max}^\mathrm{(cavity)} = \SI{3}{\mega\hertz}$, and $u_\mathrm{max}^\mathrm{(transmon)} = \SI{20}{\mega\hertz}$ were used.
For the optimizations whose results are shown in Fig.~\ref{fig:ax_shortgate} and Fig.~\ref{fig:ax_s_strongerdisplacement}, a weaker constraint ($f_\mathrm{max}=\SI{45}{\mega\hertz}$, $\Delta t=\SI{1}{\nano\second}$, $u_\mathrm{max}^\mathrm{(cavity)} = \SI{15}{\mega\hertz}$, and $u_\mathrm{max}^\mathrm{(transmon)} = \SI{20}{\mega\hertz}$) was used.

\subsection{Evaluation of Gate Error} 
\label{ax:eval_gatefid}

The quantum state dynamics with gate operations obeying the Schr\"{o}dinger or the Lindblad equation were calculated by QuTiP~\cite{qutip}.
The Lindblad equation included cavity photon loss, transmon relaxation, transmon pure dephasing, and transmon thermal excitation.
Since Eq.(\ref{eq:ax_def_fidelity}) includes an integral, we computed the gate fidelity as an average:
\begin{eqnarray}
  F\simeq \frac{1}{N_\mathrm{av}} \sum_i F \left( \hat U (|g\rangle\langle g| \otimes |\phi_i\rangle\langle\phi_i| ) \hat U^\dagger, \mathcal{E}\left(|g\rangle\langle g| \otimes |\phi_i\rangle\langle\phi_i| \right) \right), \label{eq:ax_fidelity}
\end{eqnarray}
where $|\phi_i\rangle$ corresponds to the initial state of the cavity.
$|\phi_i\rangle$ is one of the six representative points of the logical Bloch sphere,
\begin{eqnarray}
  |\phi_i\rangle \in \left\lbrace |0_L\rangle, |1_L\rangle, \frac{|0_L\rangle\pm|1_L\rangle}{\sqrt{2}}, \frac{|0_L\rangle\pm i|1_L\rangle}{\sqrt{2}} \right\rbrace.
\end{eqnarray}

\subsection{Example of Optimized Gates} 
\label{ax:example_waveform}

\begin{figure*}[htbp]
  \centering
  \includegraphics[width=.9\textwidth]{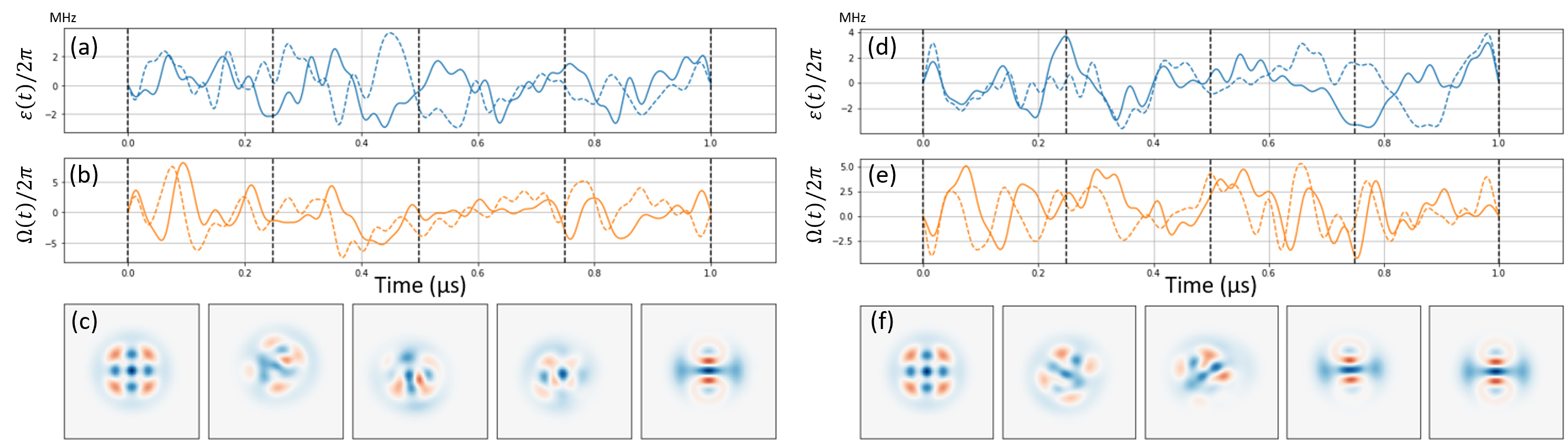}
  \caption{
    Two examples of optimized gates implementing the Hadamard gate
    The gate time is \SI{1}{\micro\second}.
    Control waveforms of cavity (a, d) and transmon (b, e).
    The solid (dashed) lines represent the in-phase (quadrature-phase) component of the waveform.
    (c, f) Wigner functions of the initial state $|0_L\rangle$, three intermediate states, and the final state.
    The timings of the intermediate states are indicated by the dashed lines.
  } \label{fig:ax_wfexamples}
\end{figure*}

Fig.~\ref{fig:ax_wfexamples} shows examples of control waveforms that implement the Hadamard gate for the Bin(1,1) code.
The waveforms were generated with different initial parameters.
Although both yield the Hadamard gate operation, their waveforms are completely different from each other, and the resultant intermediate states are also different.

%%%%%%%%%%%%%%%%%%%%%%%%%%%%%%%%%%%%%%%%%%%%%%%%%%%%%%%%%%%%%%%%%%%%%%%%%%%%%%%%
\section{Gate Error Model for Arbitrary Operations}

\subsection{Details of Gate Error Model} 
\label{ax:model}

This section derives a gate error model from the Lindblad equation.
Let $\hat H(t)$ be a time-dependent Hamiltonian realizing the desired unitary $\hat U_\mathrm{target}$.
Without decoherence, the system dynamics can be written by the Schr\"{o}dinger equation $\frac{d|\psi(t)\rangle}{dt} = -i \hat H(t) |\psi(t)\rangle$ with $|\psi(t)\rangle = \hat U_{(t, 0)} | \psi_i\rangle$, where $\hat U_{(t_1, t_0)}$ is the unitary propagator from $t_0$ to $t_1$, and $|\psi_i\rangle \left(=|\psi(0)\rangle\right)$ corresponds to the initial state of the system.
With decoherence included, the system dynamics obey the Lindblad equation $\frac{d \hat\rho(t)}{dt} = -i [\hat H(t), \hat\rho(t)] + \gamma \left( \hat L \hat\rho(t) \hat L^\dagger - \frac{1}{2}\lbrace \hat L^\dagger \hat L, \hat\rho(t) \rbrace \right) $, where $\rho(t)$, $\gamma$, and $\hat L$ respectively correspond to an intermediate state, rate of decoherence, and a jump operator representing the decoherence.
The intermediate state is $\rho(t) = \mathcal{E}_{(t, 0)} (|\psi_i\rangle\!\langle\psi_i|)$, where $\mathcal{E}_{(t_1, t_0)}$ is a quantum process from $t_0$ to $t_1$.
To consider the decoherence effect, we compare the two dynamics $\hat U_{(t + \Delta t, t)}$ and $\mathcal{E}_{(t+\Delta, t)}$ in a small time step $\Delta t$.
Starting from time $t$ and the state $|\psi(t)\rangle$, the state fidelity between the two states at $t+\Delta t$ with and without the decoherence is given by
\begin{equation}
  F_{(t+\Delta t, t)} = \langle\psi(t)|\hat U_{(t+\Delta t, t)}^\dagger \, \mathcal{E}_{(t+\Delta t, t)}\big(|\psi(t)\rangle\!\langle\psi(t)|\big) \, \hat U_{(t+\Delta t, t)} |\psi(t)\rangle .
\end{equation}
In accordance with the Schr\"{o}dinger and Lindblad equations, we can replace $\hat U_{(t+\Delta t, t)} \rightarrow 1 - i\Delta t \hat H(t) + \mathcal{O}(\Delta t^2)$ and $\mathcal{E}_{(t+\Delta t, t)} (\hat\sigma) \rightarrow \hat\sigma - i\Delta t [\hat H(t), \hat\sigma] + \Delta t \cdot \gamma \left( \hat L \hat\sigma \hat L^\dagger - \frac{1}{2}\lbrace \hat L^\dagger \hat L, \hat\sigma \rbrace \right)$, where we use $\hat\sigma = |\psi(t)\rangle\!\langle\psi(t)|$ for simplicity.
A small error induced by decoherence in the small time $\Delta t$ is expressed as
\begin{eqnarray}
  \Delta r^\prime (t;\psi_i) &= 1 - F_{(t+\Delta t, t)} \\
  &\simeq \Delta t \cdot \gamma \left( \langle\psi(t)|\hat L^\dagger \hat L|\psi(t)\rangle - \langle\psi(t)|\hat L|\psi(t)\rangle\!\langle\psi(t)|\hat L^\dagger|\psi(t)\rangle  \right) \label{eq:r_1st_approx},
\end{eqnarray}
up to first order in $\Delta t$.
Next, we define the error probability $p(t;\psi_i)$, unnormalized fidelity $l^\prime(t;\psi_i)$, and error susceptibility $s(t;\psi_i)$ at $t$ with a dependence on the initial state:
\begin{eqnarray}
  p(t;\psi_i) &= \langle\psi(t)|\hat L^\dagger \hat L|\psi(t)\rangle, \\
  l^\prime(t;\psi_i) &= \langle\psi(t)|\hat L|\psi(t)\rangle\!\langle\psi(t)|\hat L^\dagger|\psi(t)\rangle, \\
  s(t;\psi_i) &= p(t;\psi_i) - l^\prime(t;\psi_i) \label{eq:def_s_witht_ax}
\end{eqnarray}
According to the definition, $p(t;\psi_i)$ corresponds to the probability that the error occurs at $t$.
Note that $l^\prime(t;\psi_i) = |\langle\psi(t)| \hat L|\psi(t)\rangle = \langle\psi(t)|\hat L^\dagger \hat L|\psi(t)\rangle \cdot \left|\langle\psi(t)| \frac{\hat L|\psi(t)\rangle}{\| \hat L|\psi(t)\rangle \|} \right|^2 = p(t;\psi_i) \cdot |\langle\psi(t) | \psi^\prime(t)\rangle|^2$ where $|\psi^\prime(t)\rangle=\frac{\hat L|\psi(t)\rangle}{\| \hat L|\psi(t)\rangle \|}$ denotes a state vector after the error occurs at $t$.
$l^\prime(t;\psi_i)$ corresponds to the product of the error probability and the fidelity between states before and after the error occurs; thus, we refer to $l^\prime$ as the \textit{unnormalized} fidelity.

In order to encapsulate the decoherence effect, we average out the dependences on time and initial states and define the \textit{error susceptibility} of the gate by
\begin{eqnarray}
  s = \int d\psi_i \int_0^{T_\mathrm{gate}} \frac{dt}{T_\mathrm{gate}} s(t;\psi_i) \label{eq:def_s_ax},
\end{eqnarray}
where $\int d\psi_i = 1$ is an average on the logical Bloch sphere (namely, averaging $|\psi_i\rangle = |g\rangle \otimes \left( \cos\frac{\theta}{2} |0_L\rangle + e^{i\phi}  \sin\frac{\theta}{2} |1_L\rangle \right)$ where $0\le \phi \le 2\pi$ and $0 \le \theta \le \pi$).
$p$ and $l^\prime$ are defined in the same fashion.
From Eq.(\ref{eq:r_1st_approx}), (\ref{eq:def_s_witht_ax}), and (\ref{eq:def_s_ax}), the decoherence-induced gate error $r^\prime$ can be written as
\begin{eqnarray}
  r^\prime &= \int d\psi_i \int_0^{T_\mathrm{gate}} dt \cdot \gamma \, s(t;\psi_i) \\
  &= T_\mathrm{gate} \cdot \gamma \cdot s \label{eq:s_definition}.
\end{eqnarray}

This model can be easily extended to multiple decoherence cases.
By extending the Lindblad equation to a multiple decoherence case, $\mathcal{E}_{(t+\Delta t, t)} (\hat\sigma) \rightarrow \hat\sigma - i\Delta t [\hat H(t), \hat\sigma] + \Delta t \cdot \sum_k \gamma_k \left( \hat L_k \hat\sigma \hat L_k^\dagger - \frac{1}{2}\lbrace \hat L_k^\dagger \hat L_k, \hat\sigma \rbrace \right)$, Eq.(\ref{eq:s_definition}) is modified to
\begin{equation}
  r^\prime = T_\mathrm{gate} \sum_k \gamma_k s_k.
\end{equation}
Namely, the gate error is expressed as a weighted sum of all the decoherence channels.

In numerical evaluations of error susceptibility, the integral of initial states $\int d\psi_i$ is replaced to an average in the same manner as for gate errors (Eq.(\ref{eq:ax_fidelity})).

%%%%%%%%%%%%%%%%%%%%%%%%%%%%%%%%%%%%%%%%%%%%%%%%%%%%%%%%%%%%%%%%%%%%%%%%%%%%%%%%
\subsection{Comparison of the Gate Error Model and Lindblad Simulations} 
\label{ax:compare_Lindblad}

\begin{figure}[htbp]
  \centering
  \includegraphics[width=.5\textwidth]{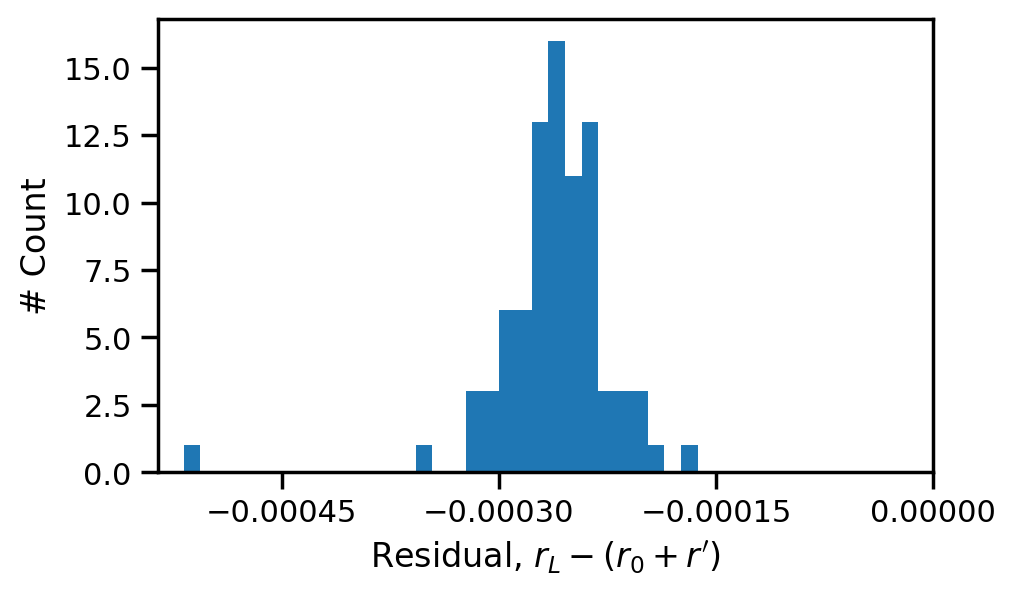}
  \caption{
    Histogram of residuals for the gate error model.
  } \label{fig:ax_compare_Lindblad}
\end{figure}

We will validate the gate error model by comparing it with the Lindblad equation for the gates investigated in Fig.~\ref{fig:susceptibility_analysis}.
Total errors $r$ can be decomposed into intrinsic errors $r_0$, which can be evaluated by solving the Schr\"{o}dinger equation, and decoherence-induced errors $r^\prime$, which can be estimated by Eq.(\ref{eq:def_rp}).
Let $r_L$ be a total error evaluated by the Lindblad equation, which will satisfy $r_L = r_0 + r^\prime$ if our error model properly captures the contributions of the given decoherences.
Fig.~\ref{fig:ax_compare_Lindblad} shows a histogram of residual $r_L - (r_0 + r^\prime)$.
Since the residuals are small, the proposed model can quantitatively capture the decoherence effects on gate errors.
Although a small bias remains, it is negligible when compared to the total errors of a few percent.
The first-order approximation in our model could yield this bias.

%%%%%%%%%%%%%%%%%%%%%%%%%%%%%%%%%%%%%%%%%%%%%%%%%%%%%%%%%%%%%%%%%%%%%%%%%%%%%%%%
\subsection{Effect of the Photon Loss Channel} 
\label{ax:toymodel}

As described in the main text, $s$ for the photon loss is close to the mean photon number $\bar{n}$ for bosonic qubits.
In this section, we prove that $s$ exactly equals $\bar{n}$ for a simpler system, a cavity mode that is driven by a displacement pulse.

Firstly, $s$ for the photon loss is conserved under displacement.
Let $|\phi\rangle$ be an arbitrary cavity state, and consider a displacement drive $|\phi\rangle \rightarrow |\phi(t)\rangle = \hat D(\epsilon t)|\phi\rangle$, where $\hat D(\alpha) = e^{\alpha \hat a^\dagger - \alpha^\ast \hat a}$.
The error susceptibility at $t$ is
\begin{equation}
  s(t;\phi) = \langle\phi|\hat D^\dagger(\epsilon t)\hat a^\dagger \hat a\hat D(\epsilon t)|\phi\rangle - |\langle\phi|\hat D^\dagger(\epsilon t)\hat a\hat D(\epsilon t)|\phi\rangle|^2.
\end{equation}
The first and second terms correspond to $p(t;\psi_i)$ and $l^\prime(t;\psi_i)$, respectively (Eq.(\ref{eq:def_s_witht_ax})).
They can be transformed as follows:
\begin{eqnarray}
  p(t;\phi) &= p(0;\phi) + \epsilon^\ast t \langle\phi|\hat a|\phi\rangle + \epsilon t \langle\phi|\hat a^\dagger|\phi\rangle + |\epsilon t|^2, \label{eq:p_disp}\\
  l^\prime(t;\phi) &= l^\prime(0;\phi) + \epsilon^\ast t \langle\phi|\hat a|\phi\rangle + \epsilon t \langle\phi|\hat a^\dagger|\phi\rangle + |\epsilon t|^2, \label{eq:lp_disp}
\end{eqnarray}
where $\hat D^\dagger(\epsilon t) \hat a \hat D(\epsilon t) = \hat a + \epsilon t$ was used, and the leftmost terms correspond to $p(0; \phi)=\langle \phi|\hat a^\dagger \hat a|\phi\rangle$ and $l^\prime(0;\phi)=\left| \langle\phi| \hat a |\phi\rangle \right|^{2}$.
The error probability $p(t;\phi)$ is increased by the displacement, but concurrently the photon distribution becomes wider, resulting in an enlarged unnormalized fidelity $l^\prime(t;\phi)$.
Each increment is given by the three rightmost terms in Eq.(\ref{eq:p_disp}) and (\ref{eq:lp_disp}), and they are identical.
Thus, the error susceptibility for the cavity photon loss is a conserved quantity under displacement drives: $s(t;\phi) = s(0;\phi)$.

Secondly, $p(0; \phi)$ equals the mean photon number of $|\phi\rangle$ because $p(0;\phi) = \langle \phi|\hat a^\dagger \hat a|\phi\rangle$.
Let us limit the initial state to the codespace: $|\phi\rangle = c_0 |0_L\rangle + c_1 |1_L\rangle$ with $|c_0|^2+|c_1|^2=1$.
Such states satisfy $p(0;\phi)=\bar{n}$ because the codewords have the same mean photon number.
The states also satisfy $l^\prime(0;\phi)=0$ because the codewords and errorwords are orthogonal: $\langle\phi| \hat a |\phi\rangle = 0$.
Thus, the error susceptibility at $t=0$ equals the mean photon number of the code: $s(0;\phi)=p(0;\phi)-l^\prime(0;\phi)=\bar{n}$.

Consequently, $s(t;\phi) = s(0;\phi) = \bar{n}$ is always satisfied for cavity modes.
Note that we considered a single harmonic oscillator, and coupling to the ancilla transmon violates the relationship.
Although it emerges as the distribution in Fig.~\ref{fig:susceptibility_analysis}(c, green), the deviations are not large for the optimized gates used in the analyses.

%%%%%%%%%%%%%%%%%%%%%%%%%%%%%%%%%%%%%%%%%%%%%%%%%%%%%%%%%%%%%%%%%%%%%%%%%%%%%%%%
\subsection{Coherence Limit of Idle Gate Error} 
\label{ax:error_transmon}

A coherence limit of an idle gate (just waiting a duration $\tau$) on a two level system is given by
\begin{eqnarray}
  r^\prime_I = \tau \cdot \left( \frac{1}{3 T_1} + \frac{1}{3 T_\phi} \right),  \label{ax:eq_idle_gate_error}
\end{eqnarray}
where $T_1$ and $T_\phi$ correspond to the relaxation time and pure dephasing time, respectively.
This equation is equivalent to Eq.(3) in Ref.~\cite{Omalley2015PRAppl} by assuming that the dephasing is like white noise.
Here, we derive this equation from our gate error model.
Let $|\psi\rangle = \cos\frac{\theta}{2} |e\rangle + e^{i\phi}\sin\frac{\theta}{2} |g\rangle$ be the initial state, and note that this state does not change in time without decoherence.

For the relaxation, the jump operator is $\hat L = \hat\sigma_-$.
The error susceptibility with state dependence is $s_\mathrm{relax}(\psi) = \langle\psi|\hat L^\dagger \hat L|\psi\rangle - |\langle\psi|\hat L|\psi\rangle|^2 = \cos^4 \frac{\theta}{2}$.
Integrating over the Bloch sphere gives the following:
\begin{equation}
  s_\mathrm{relax} = \int d\psi \cdot s_\mathrm{relax}(\psi) = \frac{1}{4\pi} \int_0^\pi \sin\theta d\theta \int_0^{2\pi} d\phi \, \cos^4\frac{\theta}{2} = \frac{1}{3}.
\end{equation}

For the pure dephasing, the jump operation is $\hat L = \frac{\hat\sigma_z}{\sqrt{2}}$.
The error susceptibility with state dependence is $s_\mathrm{dep}(\psi) = \langle\psi|\hat L^\dagger \hat L|\psi\rangle - |\langle\psi|\hat L|\psi\rangle|^2 = \frac{1}{2} - \frac{1}{2} \cos^2 \theta$.
Integrating over the Bloch sphere gives the following:
\begin{equation}
  s_\mathrm{dep} = \int d\psi \cdot s_\mathrm{dep}(\psi) = \frac{1}{4\pi} \int_0^\pi \sin\theta d\theta \int_0^{2\pi} d\phi \left( \frac{1}{2} - \frac{1}{2} \cos^2 \frac{\theta}{2} \right) = \frac{1}{3}.
\end{equation}

Thus, we have reproduced Eq.(\ref{ax:eq_idle_gate_error}) with our model.

%%%%%%%%%%%%%%%%%%%%%%%%%%%%%%%%%%%%%%%%%%%%%%%%%%%%%%%%%%%%%%%%%%%%%%%%%%%%%%%%
\subsection{Standard Deviations of Transmon Relaxation and Dephasing} 
\label{ax:std_s_transmon}

This section shows that $\mathrm{Std}[\hat\sigma_z] \ge \mathrm{Std}[\hat\sigma_z^2]$ in a simple case.
Let us assume that a transmon state $|\psi\rangle = \cos\frac{\theta}{2} |e\rangle + e^{i\phi}\sin\frac{\theta}{2} |g\rangle$ uniformly distributes on a Bloch sphere.
According to $\langle\hat\sigma_z\rangle=\cos\theta$, 
\begin{eqnarray}
  \mathrm{Ave}[\langle\hat\sigma_z\rangle] &= \frac{1}{4\pi} \int_0^\pi \sin\theta d\theta \int_0^{2\pi} d\phi \cdot \cos\theta = 0, \\
  \mathrm{Ave}[\langle\hat\sigma_z\rangle^2] &= \frac{1}{4\pi} \int_0^\pi \sin\theta d\theta \int_0^{2\pi} d\phi \cdot \cos^2\theta = \frac{1}{3}, \\
  \mathrm{Ave}[\langle\hat\sigma_z\rangle^4] &= \frac{1}{4\pi} \int_0^\pi \sin\theta d\theta \int_0^{2\pi} d\phi \cdot \cos^4\theta = \frac{1}{5}.
\end{eqnarray}
Thus we got $\mathrm{Std}[\langle\hat\sigma_z\rangle] = \sqrt{1/3 - 0^2} \simeq 0.577$ and $\mathrm{Std}[\langle\hat\sigma_z\rangle^2] = \sqrt{1/5 - (1/3)^2} \simeq 0.298$.
Although the statistical distribution of $\langle\hat\sigma_z\rangle$ was not uniform (note that the horizontal positions of the blue dots in Fig.~\ref{fig:susceptibility_analysis}(c) correspond to the average of transmon excitation: $p_\mathrm{relax} = (1 + \langle\hat\sigma_z\rangle) / 2$), this relationship could be satisfied because $-1 \le \langle\hat\sigma_z\rangle \le 1$.

\section{Gate Errors for Other Gates} 
\label{ax:comparison_gate}

\begin{figure}[htbp]
  \centering
  \includegraphics[width=.45\textwidth]{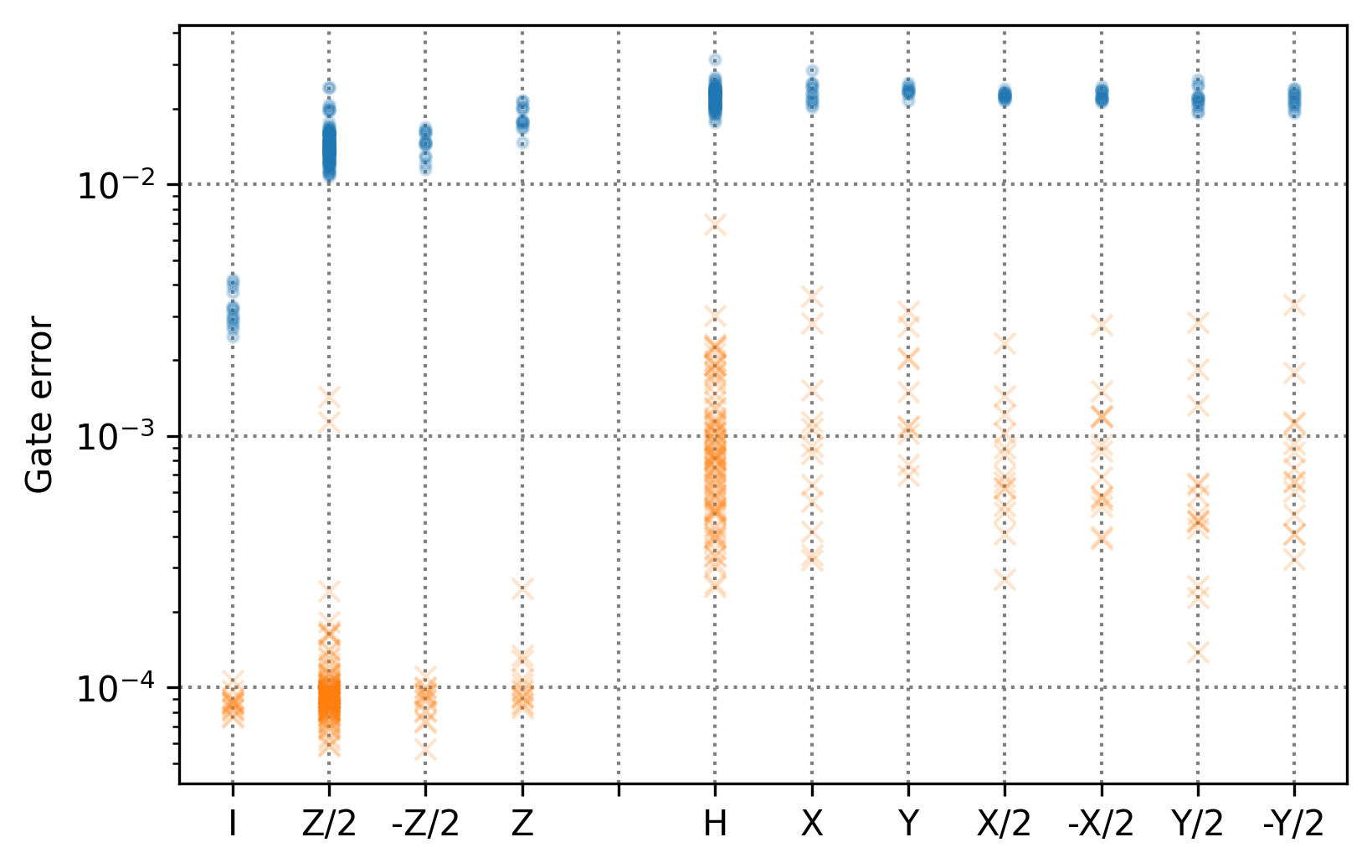}
  \caption{
    Numerically evaluated errors in various gates.
    Blue dots (orange crosses) are gate errors evaluated with (without) decoherences.
    The left four gates are phase gates that preserve the photon distribution of the cavity state; the others change the photon distribution.
  } \label{fig:ax_gatedependence}
\end{figure}

Gate operations on bosonic qubits are classified into two groups: (1) phase gates that preserve the photon distributions and (2) other gates that change the photon distribution.
Fig.~\ref{fig:ax_gatedependence} shows the gate dependence of the gate errors for the Bin(1,1) code.
It indicates that phase gates generally exhibit lower gate errors.
The gate time was fixed to \SI{1}{\micro\second}, and several gates were optimized with different initial parameters, resulting in the vertical distributions.
The decoherence rates used in this section were same to the rates used in Fig.~\ref{fig:time_vs_error}

\begin{figure*}[htbp]
 \centering
 \includegraphics[width=.9\textwidth]{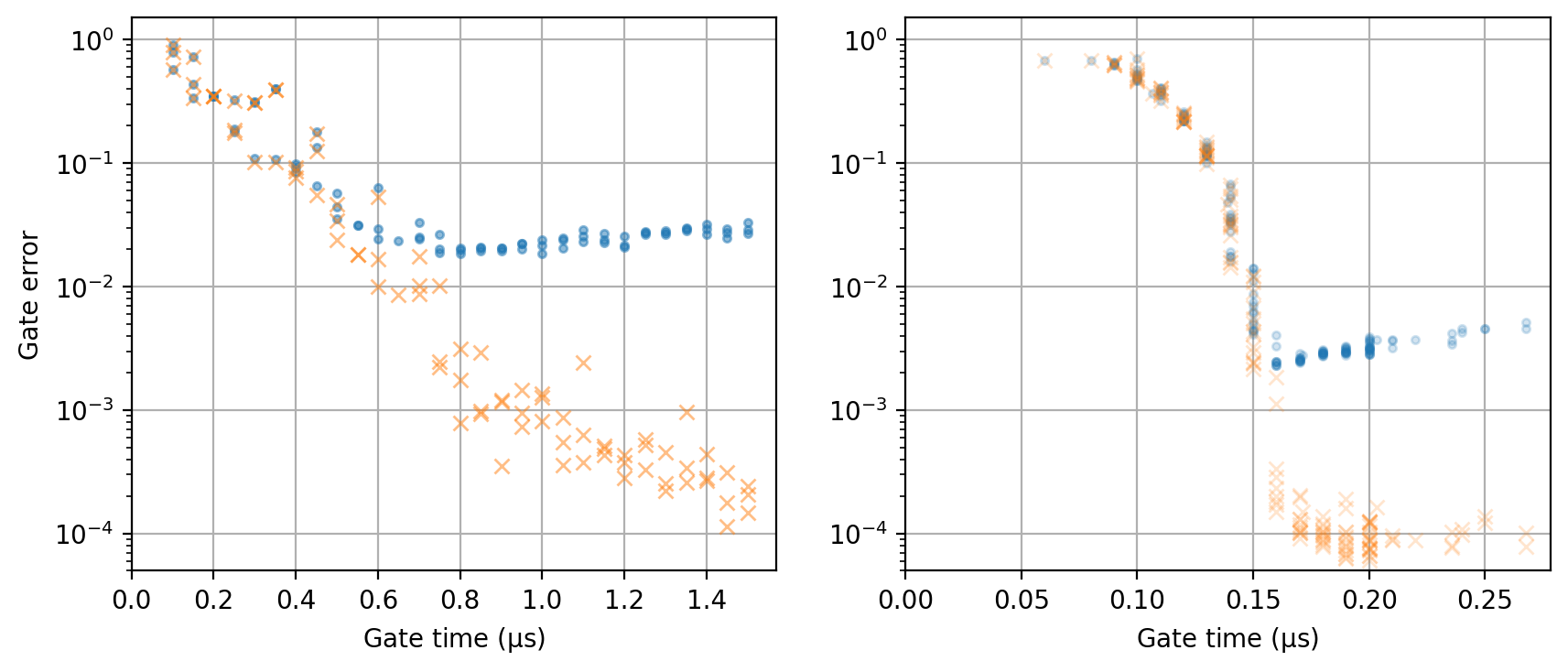}
  \caption{
    Error of the Hadamard gate (left) and $Z$ gate (right) as a function of gate time.
    Dots (crosses) are gate errors evaluated by solving the Lindblad equation with (without) decoherences.
  } \label{fig:ax_timeanderror_gatedep}
\end{figure*}

As well as the difference in errors with the specific gate time, the two groups exhibit different dependencies on gate times.
By comparing with a fixed gate time, the Hadamard gates exhibit larger errors than the $Z$ gates (Fig.~\ref{fig:ax_timeanderror_gatedep}).
Note that the saturation behavior of the $Z$ gate around $10^{-4}$ originates from the optimization terminating when the gate error fell below $10^{-4}$.

%%%%%%%%%%%%%%%%%%%%%%%%%%%%%%%%%%%%%%%%%%%%%%%%%%%%%%%%%%%%%%%%%%%%%%%%%%%%%%%%
\section{Gate Errors with Other Encodings} 
\label{ax:approxerror_othercode}

\begin{figure}[htbp]
  \centering
  \includegraphics[width=.45\textwidth]{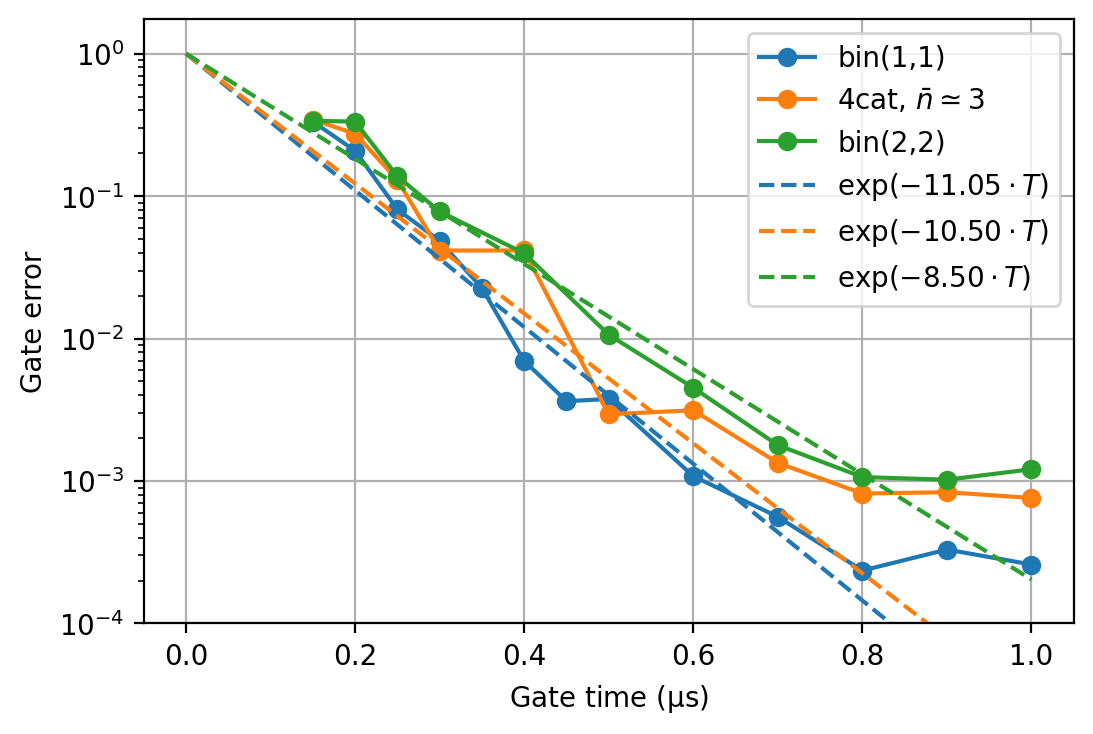}
  \caption{
    Gate time versus intrinsic gate error for several encodings.
  } \label{fig:ax_shortgate}
\end{figure}

In this section, we compare the Bin(1,1) and other encodings (the 4-leg cat and Bin(2,2), see \ref{ax:code_definition}) to provide similar analyses in Section~\ref{sec:approxerror}.
The cat size was $\alpha = \sqrt{3}$, and the mean photon number of the 4-leg cat was $\bar{n} \simeq 3$.
To obtain gate-time dependence of intrinsic gate errors, Hadamard gates for each encodings were generated by numerical optimization with several gate times (Fig.~\ref{fig:ax_shortgate}).
We fitted the gate errors by an empirical equation $r_0(T) \sim \exp(-a T)$ with $a=11.05$ for the Bin(1,1), $10.50$ for the 4-leg cat, and $8.50$ for the Bin(2,2) code.
Note that weaker constraints (\ref{ax:grape_detail}), namely a larger amplitude and wider bandwidth, were used in these optimizations.
The speed of the gate operation should be limited by the dispersive interaction when the strength and bandwidth of the waveforms are sufficient.
The aim of using weaker constraints is to ensure that intrinsic errors are limited mainly due to the dispersive interaction, not constraints in optimization.

\begin{figure}[htbp]
  \centering
  \includegraphics[width=.75\textwidth]{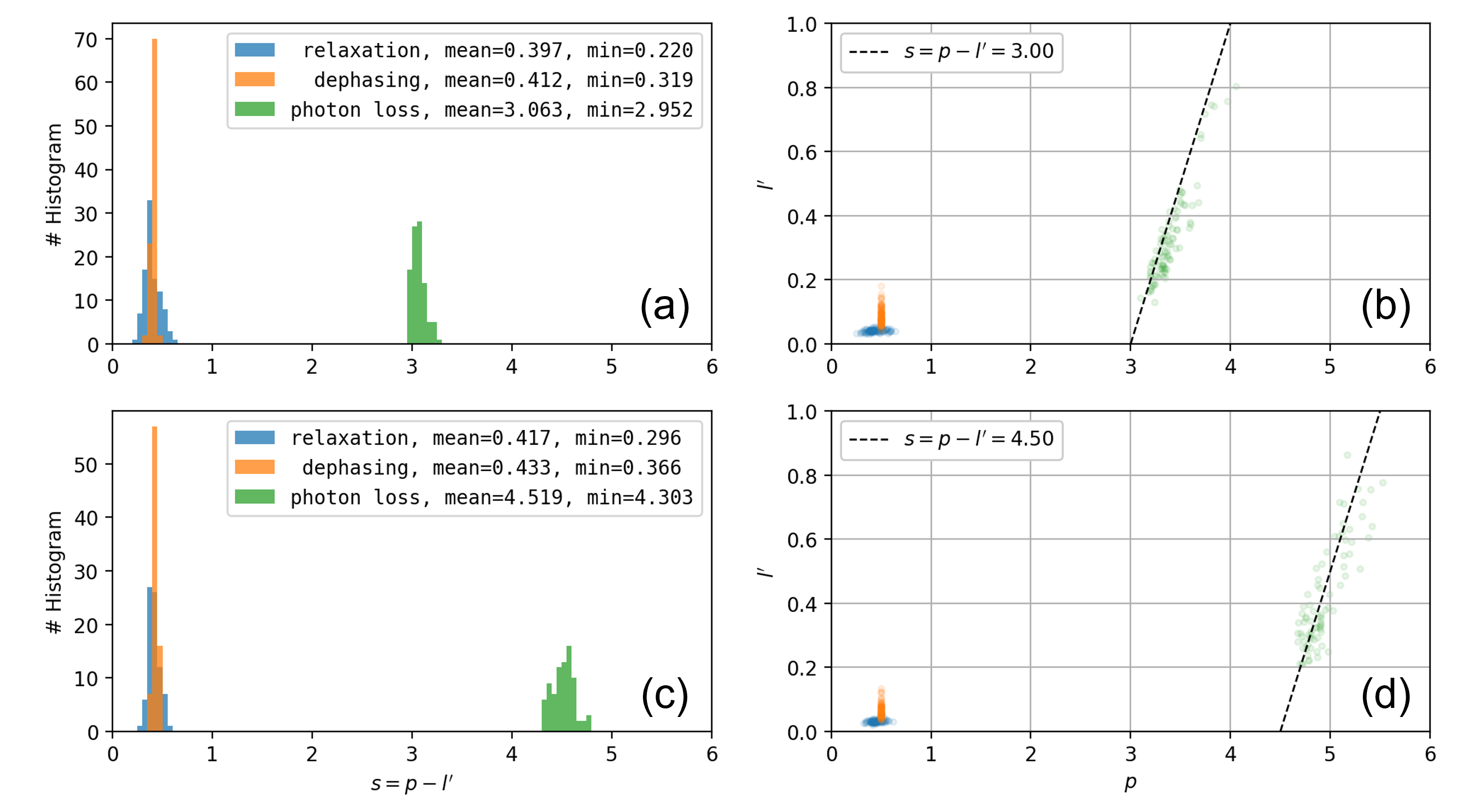}
  \caption{
    Code dependence of error susceptibility statistics:
    (a) histogram of the error susceptibility of Hadamard gates on the 4-leg cat code with $\bar{n}\simeq 3$;
    (b) relationship between $p$ and $l^\prime$;
    (c-d) those of the Bin(2,2) code.
  } \label{fig:ax_s_codedep}
\end{figure}

Fig.~\ref{fig:ax_s_codedep} shows the error susceptibilities of Hadamard gates for the 4-leg cat and Bin(2,2) code.
The gate time was fixed to \SI{1}{\micro\second}.
The codewords are different, but the error susceptibilities exhibit a similar statistics to that of the Bin(1,1) code; namely around 0.3 for transmon relaxation and dephasing, and around $\bar{n}$ for cavity photon loss.

\begin{figure}[htbp]
  \centering
  \includegraphics[width=.75\textwidth]{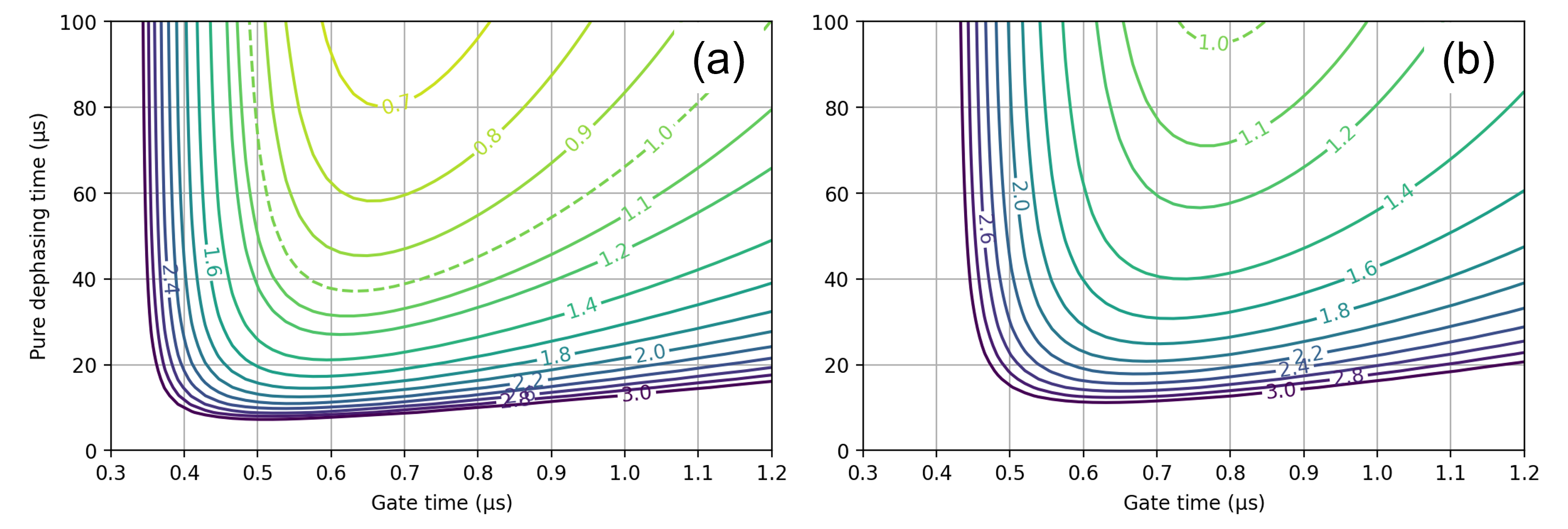}
  \caption{
    Approximate gate errors in unit of percent for (a) the 4-leg cat ($\bar{n}\simeq3$) and (b) Bin(2,2) codes.
  } \label{fig:ax_approxerror_codedep}
\end{figure}

As in Section~\ref{sec:approxerror} in the main text, we can provide rough estimations of the achievable gate errors for these encodings:
\begin{equation}
  r \ge e^{- a \cdot T_\mathrm{gate}\si{[\micro\second]} } + T_\mathrm{gate} \cdot \bigg( \frac{1}{T_1} \cdot s_1^\mathrm{(min)} + \frac{1}{T_\phi} \cdot s_\phi^\mathrm{(min)} + \kappa \cdot \frac{s_\mathrm{cav}^\mathrm{(min)}}{\bar{n}} \bar{n} \bigg).
\label{eq:ax_approx}
\end{equation}
The minimum error susceptibilities are shown in the legend of Fig.~\ref{fig:ax_s_codedep}.
Fig.~\ref{fig:ax_approxerror_codedep} shows the approximate gate errors as calculated by Eq.(\ref{eq:ax_approx}) as functions of the transmon dephasing time and gate time.


%%%%%%%%%%%%%%%%%%%%%%%%%%%%%%%%%%%%%%%%%%%%%%%%%%%%%%%%%%%%%%%%%%%%%%%%%%%%%%%%
\section{Error Susceptibility for the Recovery Gate} 
\label{ax:comparison_QEC}

\begin{figure}[htbp]
  \centering
  \includegraphics[width=.9\textwidth]{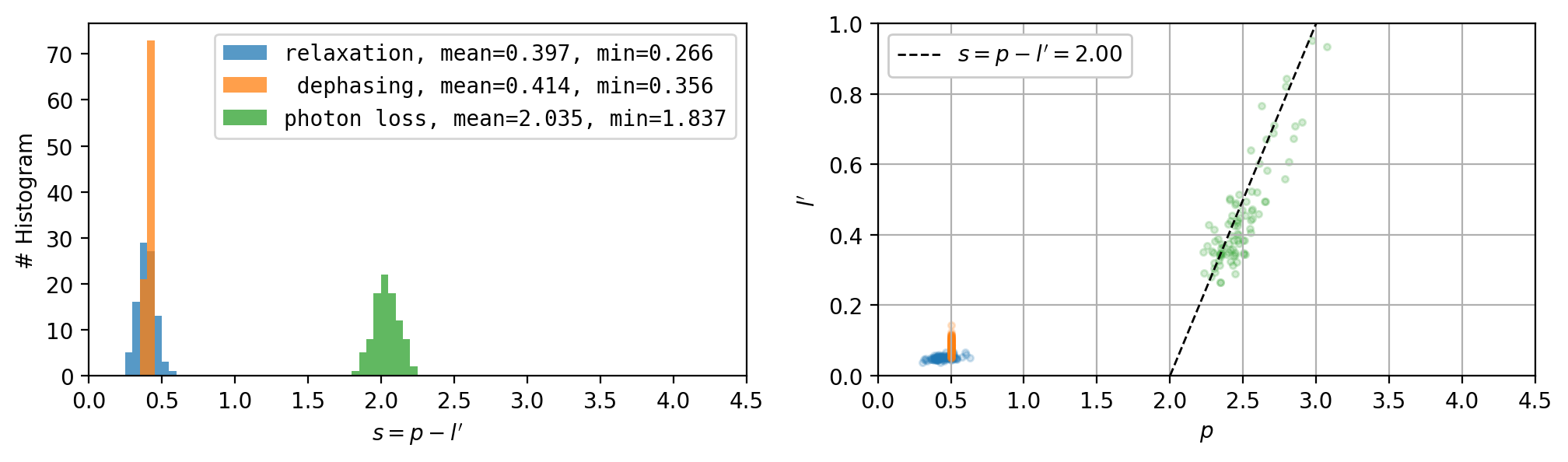}
  \caption{
    Statistics of error susceptibility for recovery gates on the Bin(1,1) code:
    (left) histogram of $s$;
    (right) relationship between $p$ and $l^\prime$
  } \label{fig:ax_s_recovery}
\end{figure}

Figure~\ref{fig:ax_s_recovery} shows the error susceptibility for recovery gates (Eq.(\ref{eq:ax_U_QEC})) for the Bin(1,1) code.
The gate time was fixed to \SI{1}{\micro\second}.
This operation is not a logical gate, but exhibits similar statistics to that of the Hadamard gate.

%%%%%%%%%%%%%%%%%%%%%%%%%%%%%%%%%%%%%%%%%%%%%%%%%%%%%%%%%%%%%%%%%%%%%%%%%%%%%%%%
\section{Error Susceptibility with Stronger Displacement} 
\label{ax:comparison_strong_displacement}

\begin{figure}[htbp]
  \centering
  \includegraphics[width=.9\textwidth]{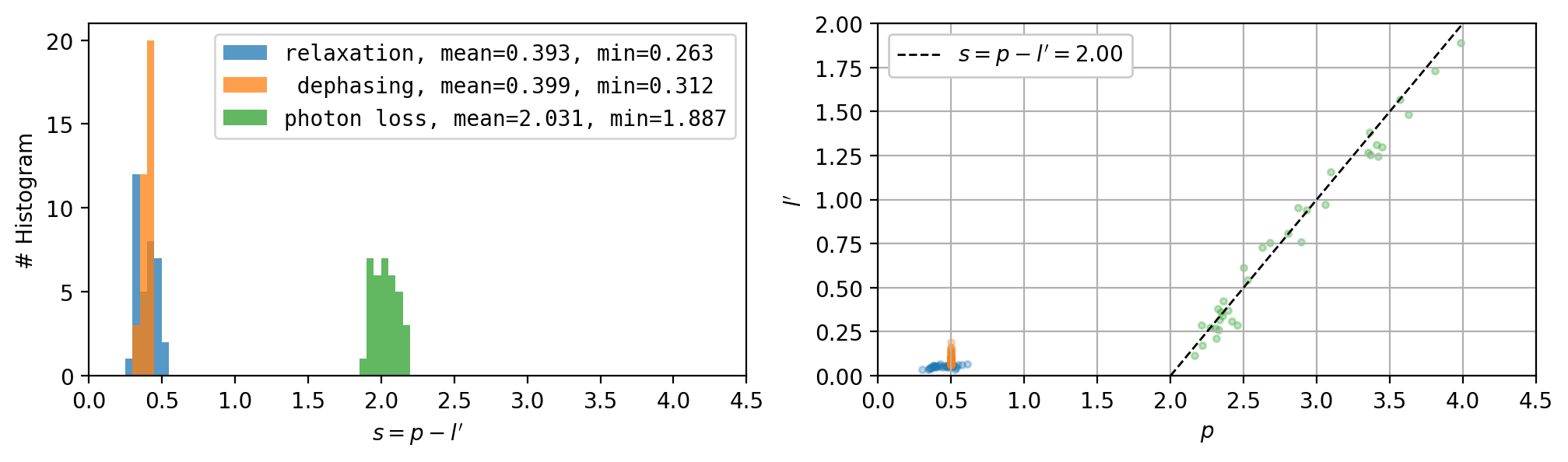}
  \caption{
    Statistics of error susceptibility for H-gates on the Bin(1,1) code with a stronger cavity displacement:
    (left) histogram of $s$;
    (right) relationship between $p$ and $l^\prime$.
  } \label{fig:ax_s_strongerdisplacement}
\end{figure}

To investigate the error susceptibility statistics for optimized gates with stronger displacements, we generated several Hadamard gates on Bin(1,1) with the weaker constraint (see \ref{ax:grape_detail}).
Fig.~\ref{fig:ax_s_strongerdisplacement} shows the resulting error susceptibilities.
The gate time was fixed to \SI{1}{\micro\second}.
Note that $p_\mathrm{loss}$ corresponds to the mean photon number during the gate, and thus the displacement strength is indicated as the horizontal position of the green dots in the right panel.
Stronger displacement pulses lead to intermediate states with higher photon number states and a higher probability of photon loss, but the figure shows that the error susceptibility is close to the mean photon number.
Thus, the contribution of the cavity photon loss channel does not depend on the displacement strength.

%%%%%%%%%%%%%%%%%%%%%%%%%%%%%%%%%%%%%%%%%%%%%%%%%%%%%%%%%%%%%%%%%%%%%%%%%%%%%%%%
\section{Gate-time Dependence of Error Susceptibility} 
\label{ax:comparison_gatetime}
\begin{figure}[htbp]
  \centering
  \includegraphics[width=.45\textwidth]{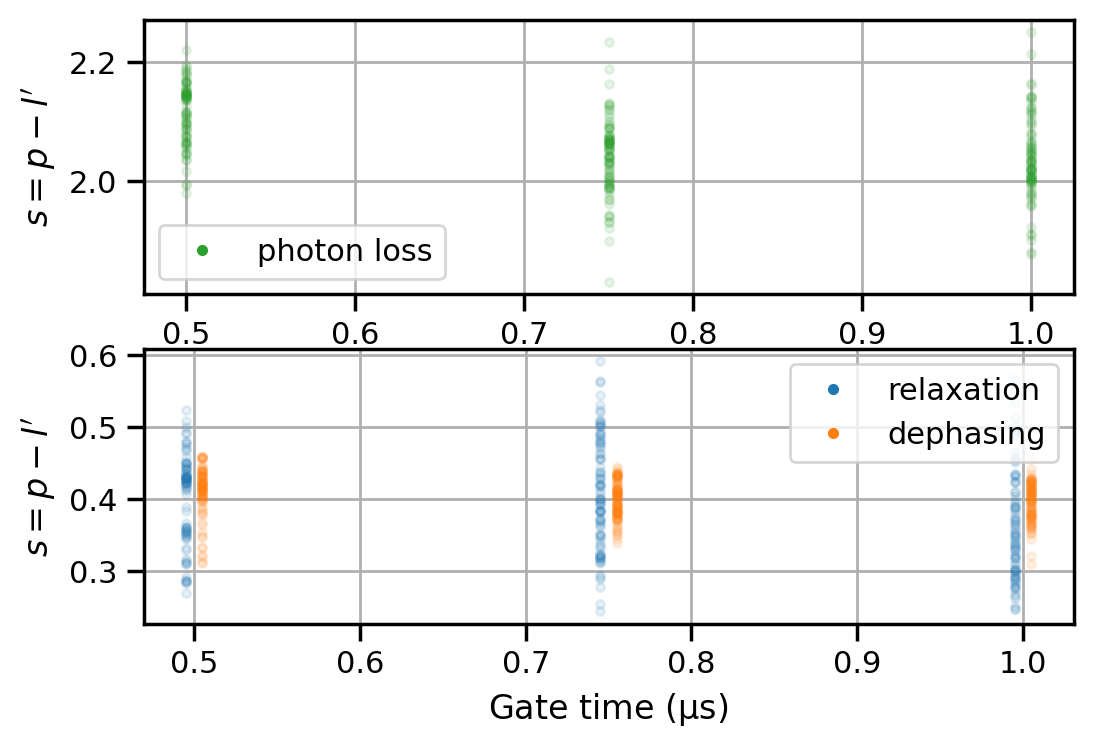}
  \caption{
    Error susceptibility of Hadamard gates with $T_\mathrm{gate}=0.5$, $0.75$, \SI{1.0}{\micro\second}.
    Upper panel: cavity photon loss, and lower panel: transmon relaxation and pure dephasing.
    The data in the lower panel is horizontally offset for clarity.
  } \label{fig:ax_s_gatetimedep}
\end{figure}

Since the error susceptibility depends on the gate time, it was further investigated for the Hadamard gate for the Bin(1,1) code (Fig.~\ref{fig:ax_s_gatetimedep}).
It was found that longer gates ($T_\mathrm{gate}=0.75$ and \SI{1.0}{\micro\second}) exhibited similar error susceptibilities while a short gate ($T_\mathrm{gate}=\SI{0.5}{\micro\second}$) exhibits a tendency to increase error susceptibility.
Decoherence-induced errors for short gates are negligible because the intrinsic errors are dominant, and thus the weak gate-time dependence of the error susceptibility can be ignored in this regime.
We assume that error susceptibility is constant in the horizontal range of Fig.~\ref{fig:ax_approxerror} in order to simplify equations like Eq.(\ref{eq:approx}).

\end{document}